\documentclass[twocolumn,preprintnumbers]{revtex4}
\usepackage{amsfonts}
\usepackage{amsmath}
\usepackage{graphicx}
\usepackage{amssymb}
\usepackage{color,soul}
\usepackage[T1]{fontenc}
\usepackage{ae,aecompl}
\usepackage{natbib}

\begin{document}

\title{Analog of a quantum heat engine using a single-spin qubit}
\author{K.~Ono$^{1,2}$ \!\footnote{E-mail address: k-ono@riken.jp}, 
	S.~N.~Shevchenko$^{3,4,5}$~\!\footnote{E-mail address: sshevchenko@ilt.kharkov.ua}, T.~Mori$^{6}$, S.~Moriyama$^{7}$%
, Franco~Nori$^{5,8}$}
\affiliation{$^1$Advanced device laboratory, RIKEN, Wako-shi, Saitama 351-0198, Japan}
\affiliation{$^{2}$CEMS, RIKEN, Wako-shi, Saitama 351-0198, Japan}
\affiliation{$^{3}$B.~Verkin Institute for Low Temperature Physics and Engineering,
Kharkov 61103, Ukraine}
\affiliation{$^{4}$V.~N.~Karazin Kharkiv National University, Kharkov 61022, Ukraine}
\affiliation{$^{5}$Theoretical Quantum Physics Laboratory, Cluster for Pioneering
Research, RIKEN, Wako-shi, Saitama 351-0198, Japan}
\affiliation{$^{6}$Device Technology Research Institute (D-Tech), National Institute of
Advanced Industrial Science and Technology (AIST), Tsukuba, Ibaraki
305-8568, Japan}
\affiliation{$^{7}$Department of Electrical and Electronic Engineering, Tokyo Denki
University, Adachi-ku, Tokyo 120-8551, Japan}
\affiliation{$^{8}$Department of Physics, The University of Michigan, Ann Arbor, MI
48109-1040, USA}

\begin{abstract}
A quantum two-level system with periodically modulated energy splitting
could provide a minimal universal quantum heat machine. We present the
experimental realization and the theoretical description of such a two-level
system as an impurity electron spin in a silicon tunnel field-effect
transistor. In the incoherent regime, the system can behave analogously to
either an Otto heat engine or a refrigerator. The coherent regime could be
described as a superposition of those two regimes, producing specific
interference fringes in the observed source-drain current.
\end{abstract}

\maketitle


\textit{Introduction.}--- Thermodynamics was originally developed for
classical many-particles systems, but recently it is being applied for the
description of individual quantum systems. This emergent field is known as
Quantum Thermodynamics~\cite{Gemmer04}. In several review articles different
types of quantum heat engines were considered, e.g. Refs.~\cite{Quan07,
Kosloff14, Goold16, Vinjanampathy16, Bhattacharjee2020}.

What essentially distinguishes a working medium of a quantum heat engine
from a classical one is the ability to be in a coherent superposition of its
states \cite{Uzdin15, Klatzow2019}. In this way, the study of
superpositional interference phenomena in prototypical quantum heat engines
is important, since this can provide a quantum advantage in their
performance, meaning the ability of a quantum heat engine to produce more
power than an equivalent classical heat engine \cite{Klatzow2019, Funo19}.

The simplest realization of a quantum heat machine would be a two-level
system. Such system, with a periodically modulated energy splitting, could
work as either a heat engine or a refrigerator, thus providing a minimal
universal quantum heat machine~\cite{Gelbwaser13, Erdman2019, Dann2020}.
Different proposed realizations are based on either natural or artificial
atoms \cite{Buluta11}, including superconducting and semiconducting
circuits, e.g., Refs.~\cite{Humphrey2002, Quan06, Abah12, Campisi13,
Campisi15, Marchegiani16, Karimi16, Erdman2017, Josefsson2018}. Recently, experimental
realizations of single-atom heat engines were demonstrated with trapped ions 
\cite{Rossnagel16, Maslennikov19, Lindenfels19} and nitrogen-vacancy centers
in diamond \cite{Klatzow2019}. In this work we present an experimental
realization and a theoretical description of heat-engine-like cycles for
highly controllable spin-1/2 states.

\begin{figure}[h]
\begin{center}
\includegraphics[width=0.5\textwidth, keepaspectratio]{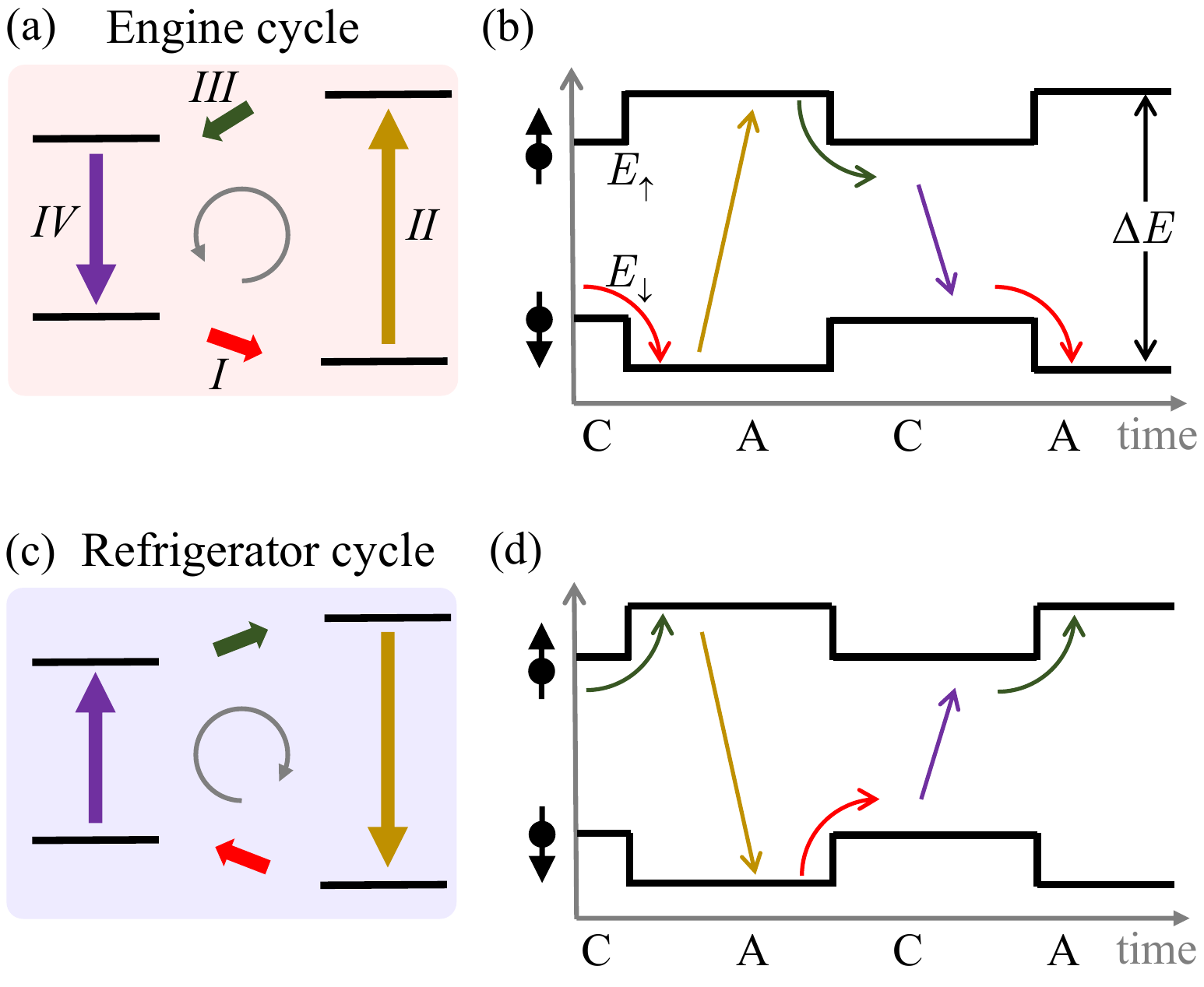}
\end{center}
\caption{Interpretation of the modulated dynamics in terms of a heat engine
or a refrigerator. The heat-engine-like cycle, shown in (a), can be realized
by modulating the energy levels, as shown in (b). The energy-level gap $%
\Delta E$ varies between larger, A, and smaller, C, values. The arrows show
the incoherent dynamics with resonant excitation during the large-energy-gap
stage A. Panels (c) and (d) present the incoherent dynamics with resonant
excitation during the small-energy-gap stage, which is reminiscent of the
refrigerator cycle.}
\label{Fig1}
\end{figure}

A heat machine can have two possible regimes, corresponding to the
engine-type and refrigerator-type cycles, as shown on the left in Fig.~\ref%
{Fig1}. The right panels of Fig.~\ref{Fig1} show their respective
realizations using a two-level system with modulated energy levels. Consider
first the situation when the period of the drive $2\pi /\Omega $ is much
larger than the decoherence time $T_{2}$, which we denote as the \textit{%
incoherent} regime. Figure~\ref{Fig1}(b)~corresponds to the situation when
the resonant driving increases the upper-level occupation during the
large-energy-gap stage [A in Fig.~\ref{Fig1}(b)], followed by a relaxation
to the ground state [C in Fig.~\ref{Fig1}(b)]. This is analogous to the
heat-engine cycle in Fig.~\ref{Fig1}(a). For another choice of parameters,
Figure~\ref{Fig1}(d)~illustrates the situation when the resonant driving
increases the upper-level occupation during the small-energy-gap stage
\textquotedblleft C\textquotedblright ,\ with the relaxation during stage
\textquotedblleft A\textquotedblright ; this is analogous to the
refrigerator cycle in Fig.~\ref{Fig1}(c).

Note the similarity between our heat-engine-like cycles and the so-called
Sisyphus lasing and cooling cycles, which also describe the periodic
evolution of a modulated dissipative two-level system, studied in Refs.~\cite%
{Grajcar08, Nori08, Skinner10, Shevchenko15, Gullans15}. These processes
take place when the period of the energy-level modulation is comparable to
the relaxation time. Then the cycle has four stages: resonant excitation
into the upper state, adiabatic evolution within this state, relaxation to
the ground state, and again adiabatic evolution.

When decreasing the ratio of the driving period, $2\pi /\Omega $, to the
decoherence time $T_{2}$, we expect that the coherence will result in
interference between the wave-function components \cite{Shevchenko10,
Forster14, Miao2019, Otxoa19}. This regime, when $T_{2}\gtrsim 2\pi /\Omega $%
, can be denoted as a \textit{coherent} regime. In this paper we study both
the incoherent quantum \textquotedblleft heat\textquotedblright\ cycles,
like in Fig.~\ref{Fig1}, and coherent dynamics, resulting in interference
fringes. We study a two-level system with modulated energy levels being
driven by a periodic frequency-modulated signal, which makes the device
highly controllable by several parameters (including the frequency and the
amplitude of modulations). We present an experimental realization and a
theoretical description of such a system, realized as an electron spin of an
impurity placed in a silicon tunnel field-effect transistor (TFET). Such
system allows a versatile control via the microwave driving and the
modulation of the Land\'{e} $g$-factor. Note that almost all works on
quantum heat engines have been theoretical. Our device has the potential to
realize these experimentally.

\textit{Device.}---Our spin-qubit device is based on a short-channel TFET.
Deep impurities are intensively ion-implanted to the channel of the TFET. At
room temperature they act as quantum dots~\cite{Ono18}. For a device with a
channel length of $\sim 80$~nm, under a source voltage $V_{\mathrm{S}}$ and
a gate voltage $V_{\mathrm{G}}$, the source-to-drain conduction is dominated
by tunneling through two impurities, i.e., one deep impurity and a shallow
impurity located near the deep impurity. For more details, see the
Supplemental Material (SM) \cite{SM}.

This double-dot-like transport exhibits spin blockade phenomena~\cite%
{ono2002current, Liu08, Giavaras13, Ono17}, and enables the time-ensemble
measure of an electron spin of one of the impurities. We focused on the
small source-drain current under the spin blockade condition. Under
appropriate dc and ac magnetic fields, the source-drain current increases
due to the lifting of the spin blockade by the electron spin resonance (ESR)
of one of the impurities~\cite{koppens2006driven}. Thus sweeping the
frequency of the ac component of the magnetic field, for a fixed dc magnetic
field, the source-drain current shows a peak at the ESR frequency, and the
peak's height is proportional to the excitation probability of the spin
qubit.

We measure the spin-blockade source-drain current of the device at $1.6$~K
with dc magnetic field $B\approx 0.28$~T and ac magnetic field with
microwave (MW) frequency $\sim 9$~GHz. An observed linewidth of $\sim 4$%
~MHz, as well as a Rabi oscillation measurement~\cite{Ono18}, show a
coherence time\ of $0.25$~$\mu $s. This timescale includes the effect of
finite lifetime of the spin on the impurity. This finite lifetime is due to
the finite mean-stay time of the electron on the impurity in the tunneling
transport through the impurities. Thus the spin on the impurity is replaced
at this timescale. Theoretically, this situation is described by introducing
the phenomenological relaxation and decoherence times: $T_{1,2}$. For a
two-level system, $T_{1}$ describes the relaxation from the excited state,
while $T_{2}$ describes the lifetime of the coherence in the system.

\textit{Qubit energy modulation.}---We found that the $g$-factor of one of
the impurities can be tuned by the gate voltage $V_{\mathrm{G}}$. This is
due to the Stark effect~\cite{Ono18b}. In our device $\sim 1\%$ of the $g$%
-factor can be tuned by changing $V_{\mathrm{G}}$ within $\pm 20$~mV. Note
that the spin-blockade condition is still kept in our device if we change $%
V_{\mathrm{G}}$ by such amount. The fast modulation of $V_{\mathrm{G}}$, and
thus the energy of the spin qubit, are carried out by adding the radio
frequency modulation to the gate electrode. Square waves with frequencies
from $0.05$ to $10$~MHz are used. See Refs.~\cite{Ono18, Ono18b} for further
details on the device, spin qubit measure, and $V_{\mathrm{G}}$ modulation;
and see Refs.~\cite{Deng13, Funo17, Funo19} for the study of other
modulation pulses; in particular, the ones realizing the counter-diabatic
driving.

\textit{MW frequency modulation.}---We use the frequency modulation (FM)
function of our MW generator, where the FM is proportional to the voltage
signal fed to the external input. The two-channel arbitrary waveform
generator was used for the $V_{\mathrm{G}}$ modulation, and the voltage was
used for microwave frequency modulation. It is important to note that the
bandwidth of the arbitrary waveform generator is $\sim 100$~MHz, so this
modulation never excites higher-lying spin states, including the $1/2$-spin
of another impurity whose ESR frequency is $1$~GHz higher than the focused
spin. Thus, changes of both the qubit energy and MW frequency can be
regarded as adiabatic. For the synchronized modulation of both the qubit
energy and the MW frequency, we used two square-wave signals with tunable
amplitudes and phase difference, and fed one signal to the gate electrode
and another one to the input for the FM signal on the MW generator.

\textit{Theoretical description.}--- We describe this spin-qubit device as a
driven two-level system, which is amplitude- and frequency- modulated, with
the pseudo-spin Hamiltonian $H(t)=B_{z}(t)\sigma _{z}/2+B_{x}(t)\sigma
_{x}/2 $. In other words, we consider a single $1/2$-spin subject to a fast
microwave driving and a slow rf modulation of both amplitude \textit{and}
frequency. Note that in our previous work~\cite{Ono18b} we only considered
amplitude modulation.

\begin{figure}[t]
\begin{center}
\includegraphics[width=0.35\textwidth, keepaspectratio]{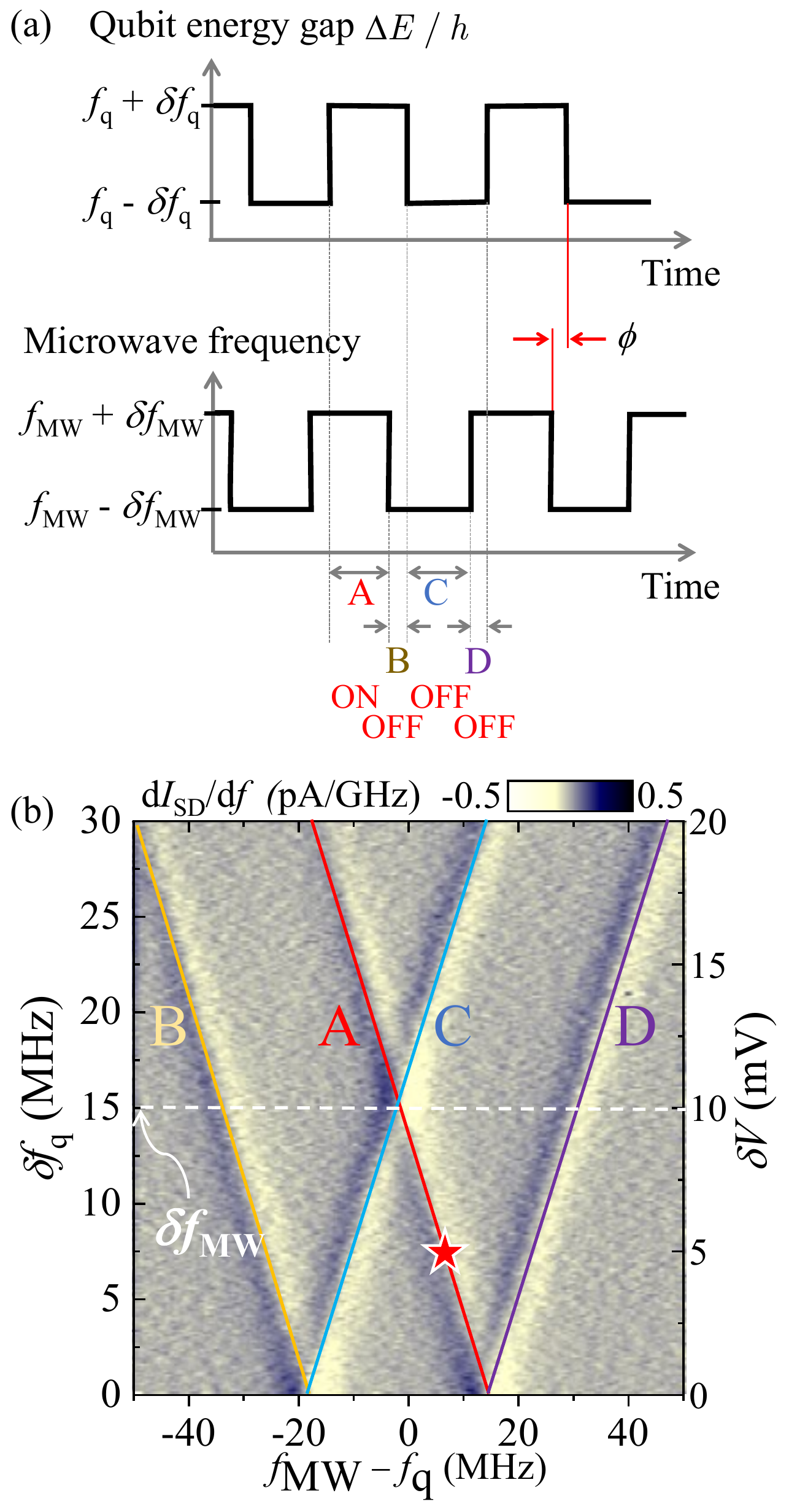}
\end{center}
\caption{Amplitude- and frequency-modulated two-level system. In the upper
panel of (a) the qubit energy $\Delta E(t)=E_{\uparrow }-E_{\downarrow }=h%
\left[ f_{\mathrm{q}}+\protect\delta \!f_{\mathrm{q}}\,s(t)\right] $ is
shown to be modulated in amplitude; in the lower panel, the modulated
driving (microwave) frequency $f_{\mathrm{FM}}(t)=f_{\mathrm{MW}}+\protect%
\delta \!f_{\mathrm{MW}}\,s_{\protect\phi }(t)$ is presented. The phase
shift $\protect\phi $ corresponds to four stages of evolution, denoted from
A to D. We show the situation when the qubit is resonantly excited during
the stage \textquotedblleft A\textquotedblright , as in Fig. \protect\ref%
{Fig1}(b). Then the coupling between the drive and the qubit is
\textquotedblleft ON\textquotedblright ~during this stage and
\textquotedblleft OFF\textquotedblright ~during the other stages. The four
possible resonant excitations are shown by the inclined lines in (b), where
the colors correspond to the four stages in (a). Panel (b) presents the
derivative of the source-drain current as a function of the frequency
detuning and the amplitude of the energy-level modulation at $\protect\delta %
\!f_{\mathrm{MW}}=15$~MHz.}
\label{Fig2}
\end{figure}

\begin{figure}[t]
\begin{center}
\includegraphics[width= 1.05\columnwidth, keepaspectratio]{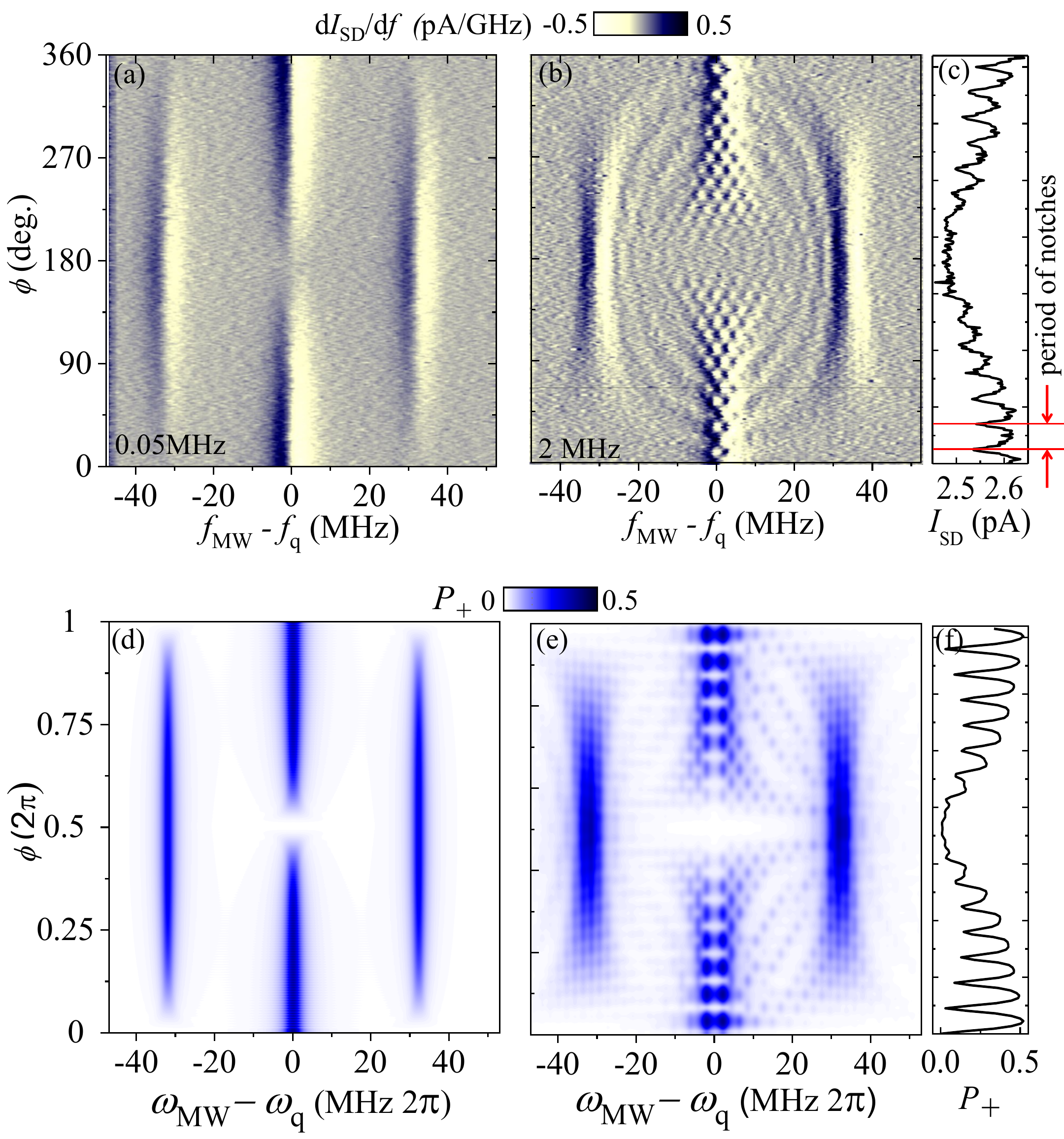}
\end{center}
\caption{Incoherent and coherent regimes at $\protect\delta \!f_{\mathrm{MW}%
}=\protect\delta \!f_{\mathrm{q}}$. Source-drain current $I_{\mathrm{SD}}$
as a function of the frequency detuning $\left( f_{\mathrm{MW}}-f_{\mathrm{q}%
}\right) $ and the phase difference $\protect\phi $, in the incoherent (a)
and coherent (b) regimes, for the modulation frequencies $\Omega /2\protect%
\pi =0.05$ and~$2$~MHz, respectively. (c) is the cross-section of (b) along
the line $f_{\mathrm{MW}}=f_{\mathrm{q}}$. (d)-(f) show the respective
upper-level occupation probability. Note the small \textquotedblleft
notches\textquotedblright\ in (b,c) and (e,f).}
\label{Fig3}
\end{figure}

The longitudinal part of the\ Hamiltonian $H(t)$ is defined by the Zeeman
splitting, $B_{z}(t)=g(t)\mu _{\mathrm{B}}B$. The time-dependent gate
voltage changes the $g$-factor by a small value \cite{Giavaras19} and we
have $B_{z}/\hbar =\omega _{\mathrm{q}}+\delta \omega _{\mathrm{q}}\,s(t)$,
with $\delta \omega _{\mathrm{q}}\ll \omega _{\mathrm{q}}$, where $\omega _{%
\mathrm{q}}=2\pi f_{\mathrm{q}}$ represents the ESR frequency, and $\delta
\omega _{\mathrm{q}}$ describes the \textit{amplitude modulation}. In this
work, we consider a square-wave modulation $s(t)=\mathrm{sgn}\left[ \cos
\Omega t\right] $. The transverse part of the Hamiltonian is defined by the
frequency-modulated MW voltage applied to the substrate, $B_{x}/\hbar
=2G\cos \left[ \omega _{\mathrm{FM}}(t)t\right] $, with a frequency $\omega
_{\mathrm{FM}}(t)=\omega _{\mathrm{MW}}+\delta \omega _{\mathrm{MW}%
}\,s_{\phi }(t)$, which is modulated by the phase-shifted signal $s_{\phi
}(t)=\mathrm{sgn}\left[ \cos \left( \Omega t+\phi \right) \right] $. Here $G$
stands for the amplitude, which is defined by the microwave power at the
microwave-generator output and $\omega _{\mathrm{MW}}=2\pi f_{\mathrm{MW}}$
is the microwave circular frequency. The modulation is assumed to be slow,
i.e. $\Omega \ll \omega _{\mathrm{MW}}$, and with a small amplitude, $\delta
\omega _{\mathrm{MW}}\ll \omega _{\mathrm{MW}}$, where $\delta \omega _{%
\mathrm{MW}}$ describes the \textit{frequency modulation}. (About amplitude
and frequency modulation, see also in Refs.~\cite{Savelev04, Savelev04b,
Ooi07, Satanin14, Silveri17, Mondal2020}.) With the Hamiltonian~$H(t)$ we
solve the Bloch equations, as described in detail in the SM~\cite{SM}, see
also Refs.~\cite{Silveri15, Shevchenko19}. Analytical stationary solutions
of the Bloch equations allow us to obtain the upper-level occupation
probability $P_{+}$, as shown in the lower part of Fig.~\ref{Fig3}, while
the numerical solution gives the time-dependent occupation $P_{+}(t)$~\cite%
{SM}.

The modulated qubit energy levels $E_{\uparrow ,\downarrow }=\pm \frac{1}{2}%
\Delta E=\pm \frac{\hbar }{2}\left[ \omega _{\mathrm{q}}+\delta \omega _{%
\mathrm{q}}\,s(t)\right] $ are plotted as a function of the dimensionless
time $\tau =\Omega t/2\pi $ in the upper panel in Fig.~\ref{Fig2}(a). The
lower panel of Fig.~\ref{Fig2}(a) presents the time dependence of the
modulated microwave frequency $\omega _{\mathrm{FM}}(t)$. This frequency is
phase shifted with respect to the energy-level modulation.

The qubit experiences a resonant excitation when $\Delta E(t)=\hbar \omega _{%
\mathrm{FM}}(t)$. This relation, written as $\omega _{\mathrm{q}}+\delta
\omega _{\mathrm{q}}\,s(t)=\omega _{\mathrm{MW}}+\delta \omega _{\mathrm{MW}%
}\,s_{\phi }(t)$, allows four possibilities for the resonant excitation: $%
\Delta \omega \equiv \omega _{\mathrm{q}}-\omega _{\mathrm{MW}}=\Delta
\omega ^{(A,B,C,D)}$, where $\Delta \omega ^{(A,B)}=-\delta \omega _{\mathrm{%
q}}\pm \delta \omega _{\mathrm{MW}}$ and $\Delta \omega ^{(C,D)}=\delta
\omega _{\mathrm{q}}\mp \delta \omega _{\mathrm{MW}}$. This means that, if
one of the conditions is met, $\Delta \omega =\Delta \omega ^{(i)}$, then
during the $i$-th stage the resonant condition is fulfilled.

Note that the above conditions are valid for relatively small modulation
frequencies, when the respective period is much larger than the decoherence
time, $2\pi /\Omega \gg T_{2}$. In this case, the incoherent dynamics during
one of the stages does not influence the dynamics of the later stages. And
indeed, in the source-drain current the four resonance conditions were
observed as the inclined lines along $\Delta \omega =\Delta \omega
^{(A...D)} $, Fig.~\ref{Fig2}(b).

Specifically, we show schematically the resonant excitation in Fig.~\ref%
{Fig2}(a) and Fig.~\ref{Fig1}(b) for the situation when the resonant
condition is for the stage \textquotedblleft $A$\textquotedblright , where $%
\Delta \omega =\Delta \omega ^{(A)}$. In Fig.~\ref{Fig1}(b) we start from
the system in the ground state; then the qubit becomes partially excited;
and later on we have full relaxation back to the ground state. This is shown
for the incoherent regime. In the coherent regime, the relaxation time of
the population is longer than the driving period (so that $T_{2}\gtrsim 2\pi
/\Omega $) and the interference between different stages takes place, which
we will consider later in more detail.

\textit{Analogy to a heat engine.}--- First, it is possible to introduce an
(effective) temperature $T$ as the value defining the energy-level
populations $P_{+}$ and $P_{-}=1-P_{+}$ with the relation $\frac{P_{-}}{P_{+}%
}=\exp \left( \frac{\Delta E}{k_{\mathrm{B}}T}\right) $ \cite{Henrich07,
Barontini2019}. Then the driven and relaxed stages, with $P_{+}$ close to $%
1/2$ and $0$ respectively, determine the cold and hot reservoir temperatures~%
\cite{Lindenfels19}. The emulation of a quantum heat engine is then
completed by associating the energy-level distance $\Delta E$ with the
volume of a working gas in the corresponding macroscopic engine~\cite%
{Henrich07, Lindenfels19}. With these, we are prepared to describe the
driven evolution of our single-spin quantum machine.

We now consider, for simplicity, $\phi =0$, while a non-zero phase shift $%
\phi $ gives an additional degree of control. Then the stages $D$ and $B$ in
Fig.~\ref{Fig2}(a) collapse and we have the alternation of the stages $A$
and $C$ only, as in Fig.~\ref{Fig1}. We start in Fig.~\ref{Fig1}(a,b) with $%
\Delta \omega =\Delta \omega ^{(A)}$. Then the changes in the energy-level
populations are shown by the four arrows, with the resonant excitation to
the upper level shown by the dark yellow arrow and relaxation to the ground
state shown by the violet arrow. Such evolution is equivalently drawn in
Fig.~\ref{Fig1}(a), which behaves like a four-stroke quantum Otto engine~%
\cite{Goold16}. During the stage $I$, the energy difference $\Delta E$
increases, corresponding to an expansion. The resonant drive during the
stage $I\!I$ increases the upper-level population $P_{+}$, as if in contact
with a hot thermal reservoir. Then we have the compression during the stage $%
I\!I\!I$. Finally, the relaxation during the stage $I\!V$ plays the role of
contacting with a cold thermal reservoir, with decreasing $P_{+}$.

An analogous evolution with the resonant excitation in the stage
\textquotedblleft $C$\textquotedblright\ is presented in Fig.~\ref{Fig1}%
(c,d). Such cycle, shown in Fig.~\ref{Fig1}(c), is reminiscent of a
four-stroke refrigerator.

We emphasize that the situation shown in Fig.~\ref{Fig1} is for the fast
relaxation case, $2\pi /\Omega \gg T_{1,2}$. If the modulating frequency $%
\Omega $ is increased, then these two values become comparable, $2\pi
/\Omega \sim T_{1,2}$. Then the interference between the different stages
can be considered as a superposition of the \textquotedblleft
heat-engine\textquotedblright\ and \textquotedblleft
refrigerator\textquotedblright\ cycles. In the measured source-drain
current, such superposition results in interference fringes, which we
present later. 

Note that the analogy with a \textquotedblleft heat
engine\textquotedblright\ is a partial one. Differences include the lack of
high- and low-temperature reservoirs. In our system, the source and the drain are tools to probe the qubit, not heat reservoirs. Namely, the electrodes are reservoirs of electrons, and are not directly related to thermal reservoirs.  A hot (thermodynamic) reservoir has more energy than a cold (thermodynamic) reservoir.  An electronic analog of this would be this: a higher-voltage (electron) reservoir has more energy than a lower-voltage (electron) reservoir.

\textit{Slow-modulation regime.}--- The measured source-drain current $I_{%
\mathrm{SD}}$ is shown in Fig.~\ref{Fig2}(b) as a function of the driving
microwave frequency $f_{\mathrm{MW}}$ and the amplitude of the energy-level
modulation $\delta \!f_{\mathrm{q}}$ for a low-frequency ($\Omega /2\pi
=0.05 $~MHz) square modulation, and the phase difference of two modulations $%
\phi =90$~degrees. The four lines of the ESR peaks are seen in Fig.~\ref%
{Fig2}(b) at such value of $\phi $. Next, we vary the phase difference $\phi 
$ between the two square-wave modulations, shown in Fig.~\ref{Fig3}(a).
While fixing the two amplitudes of the square modulations to $\delta \!f_{%
\mathrm{MW}}=\delta \!f_{\mathrm{q}}$, the phase difference $\phi $ is
changed from $0$ to $360$~degrees. The heights of the three ESR peaks are
changed as in Fig.~\ref{Fig3}(a). At $\phi =0$, the height of the center
peak is maximum and two side peaks disappear. The condition with $\phi =0$
and $f_{\mathrm{MW}}-f_{\mathrm{q}}=0$ can be called \textit{in-phase},
where the phase of the two modulations matches and the qubit is always in
resonance with the microwaves. In contrast, at $\phi =180$~deg., the center
peak disappears and the heights of two side peaks have their maximum. At
this \textit{out-of-phase} condition, $\phi =180$~deg. and $f_{\mathrm{MW}%
}-f_{\mathrm{q}}=0$, the qubit is always driven out of resonance. The height
of the center ridge evolves linearly from its maximum (at the in-phase
condition) to zero (at the out-of-phase condition), Fig.~\ref{Fig3}(a). The
height simply reflects the duration of the \textquotedblleft
ON\textquotedblright\ stage, when the microwave is in resonance with the
qubit, see Fig.~\ref{Fig2}(a).

\textit{Fast modulation: interference pattern.}--- Now we increase the
modulation frequency up to around $T_{2}{}^{-1}$. Figure~\ref{Fig3}(b) shows
an intensity plot similar to Fig.~\ref{Fig3}(a), but now the modulation
frequency $\Omega $ is set to $2$~MHz$\cdot 2\pi $. In Figure~\ref{Fig3}(b)
there are three vertical ridges, and their positions and heights are similar
to the case of the slow modulation. However, there are many fringes in the
interference pattern. A ripple-like pattern is seen between the center and
side ridges. The edges of the ridges show fan-shaped broadenings, and,
importantly, fine periodic interference patterns appear only around the
central ridge. Similar to the slow modulation, the central ridge height is
maximal at the in-phase condition and zero at the out-of-phase condition.
However, due to this interference pattern, the $\phi $ dependence of the
central ridge height shows a \textquotedblleft beating\textquotedblright\
pattern with periodically appearing notches [a cross-section shown in Fig.~%
\ref{Fig3}(c)].

We repeat similar measurements with various modulation frequencies,
modulation amplitudes, and MW power \cite{SM}. For each setting, we repeat
similar calibrations as in Fig.~\ref{Fig3}, and make sure that the condition 
$\delta \!f_{\mathrm{MW}}=\delta \!f_{\mathrm{q}}$ is met. We focus on the
period of the notches observed on the central ridge, as indicated in Fig.~%
\ref{Fig3}(c), as a characteristic parameter of the interference. The
results of calculations are presented in Fig.~\ref{Fig3}(d-f). The period of
the notches is proportional to the modulation frequency $\Omega $, i.e. $%
\Delta \phi /2\pi =\Omega /\delta \omega _{\mathrm{MW}}$, inversely
proportional to the modulation amplitude $\delta \omega _{\mathrm{q}}$, i.e. 
$\Delta \phi /2\pi =\Omega /\delta \omega _{\mathrm{q}}$, and independent of
the microwave power $G$ \cite{SM}.

\textit{Conclusion.}--- We presented a detailed study of an energy- and
frequency-modulated two-level system. The experimental realization of this
was a spin-qubit device, based on a deep impurity in a short-channel silicon
TFET. This was shown to work analogously to a heat engine or a refrigerator
in the incoherent regime, at slow energy-gap modulation; and displaying
interference fringes in the coherent regime, when the modulation period
becomes larger than the decoherence time. Note that the coexistence of a
classical heat pump and a refrigerator would functionally cancel each other;
whereas the quantum superposition of an engine and a cooler exhibits novel
interferometric effects. Due to such interference, the quantum thermodynamic
system can quickly switch its function between engine or refrigerator
regimes, and respond fast to external signals, which is not possible with
classical systems. Our impurity-based spin-qubit system has a set of
parameters which reliably control its state. This makes it useful for
possible applications, such as a future universal quantum heat engine.

\vspace*{-1cm}  

\begin{acknowledgments}
	
	\vspace*{-0.5cm}  
	
We thank Ken Funo, Neill Lambert, and M.~Fernando Gonzalez-Zalba for
discussions. This work was supported by JST CREST JPMJCR1871, MEXT Q-LEAP
JPMXS0118069228, JSPS KAKENHI 15H04000, 17H01276. F.N. is supported in part
by: NTT Research, Army Research Office (ARO) (Grant No.~W911NF-18-1-0358),
Japan Science and Technology Agency (JST) (via Q-LEAP and the CREST Grant
No.~JPMJCR1676), Japan Society for the Promotion of Science (JSPS) (via the
KAKENHI Grant No.~JP20H00134, and the grant JSPS-RFBR Grant
No.~JPJSBP120194828), and the Grant No.~FQXi-IAF19-06 from the Foundational
Questions Institute Fund (FQXi), a donor advised fund of the Silicon Valley
Community Foundation. Research of S.N.S. was sponsored by the Army Research
Office and was accomplished under Grant Number W911NF-20-1-0261.
\end{acknowledgments}

\vspace*{-0.7cm}  

\nocite{apsrev41Control} 
\bibliographystyle{apsrev4-1}
\bibliography{references}

\end{document}


\title{Supplemental Material for:\\
Analog of a quantum heat engine using a single-spin qubit}
\author{K.~Ono$^{1,2}$\thanks{%
E-mail address:k-ono@riken.jp}, S.~N.~Shevchenko$^{3,4,5}$, T.~Mori$^{6}$,
S.~Moriyama$^{7}$, Franco~Nori$^{5,8}$}
\affiliation{$^1$Advanced device laboratory, RIKEN, Wako-shi, Saitama 351-0198, Japan}
\affiliation{$^{2}$CEMS, RIKEN, Wako-shi, Saitama 351-0198, Japan}
\affiliation{$^{3}$B.~Verkin Institute for Low Temperature Physics and Engineering,
Kharkov 61103, Ukraine}
\affiliation{$^{4}$V.~N.~Karazin Kharkov National University, Kharkov 61022, Ukraine}
\affiliation{$^{5}$Theoretical Quantum Physics Laboratory, Cluster for Pioneering
Research, RIKEN, Wako-shi, Saitama 351-0198, Japan}
\affiliation{$^{6}$Device Technology Research Institute (D-Tech), National Institute of
Advanced Industrial Science and Technology (AIST), Tsukuba, Ibaraki
305-8568, Japan}
\affiliation{$^{7}$Department of Electrical and Electronic Engineering, Tokyo Denki
University, Adachi-ku, Tokyo 120-8551, Japan}
\affiliation{$^{8}$Department of Physics, The University of Michigan, Ann Arbor, MI
48109-1040, USA}
\date{\today }

\begin{abstract}
In this Supplemental Material (SM) we present details of the device,
experiment, and theory.
\end{abstract}

\pacs{73.63.Kv, 73.23.Hk, 76.30.-v}
\maketitle

\section{Impurity-based spin qubit}

\begin{figure}[p]
\par
\includegraphics[width=15.5cm]{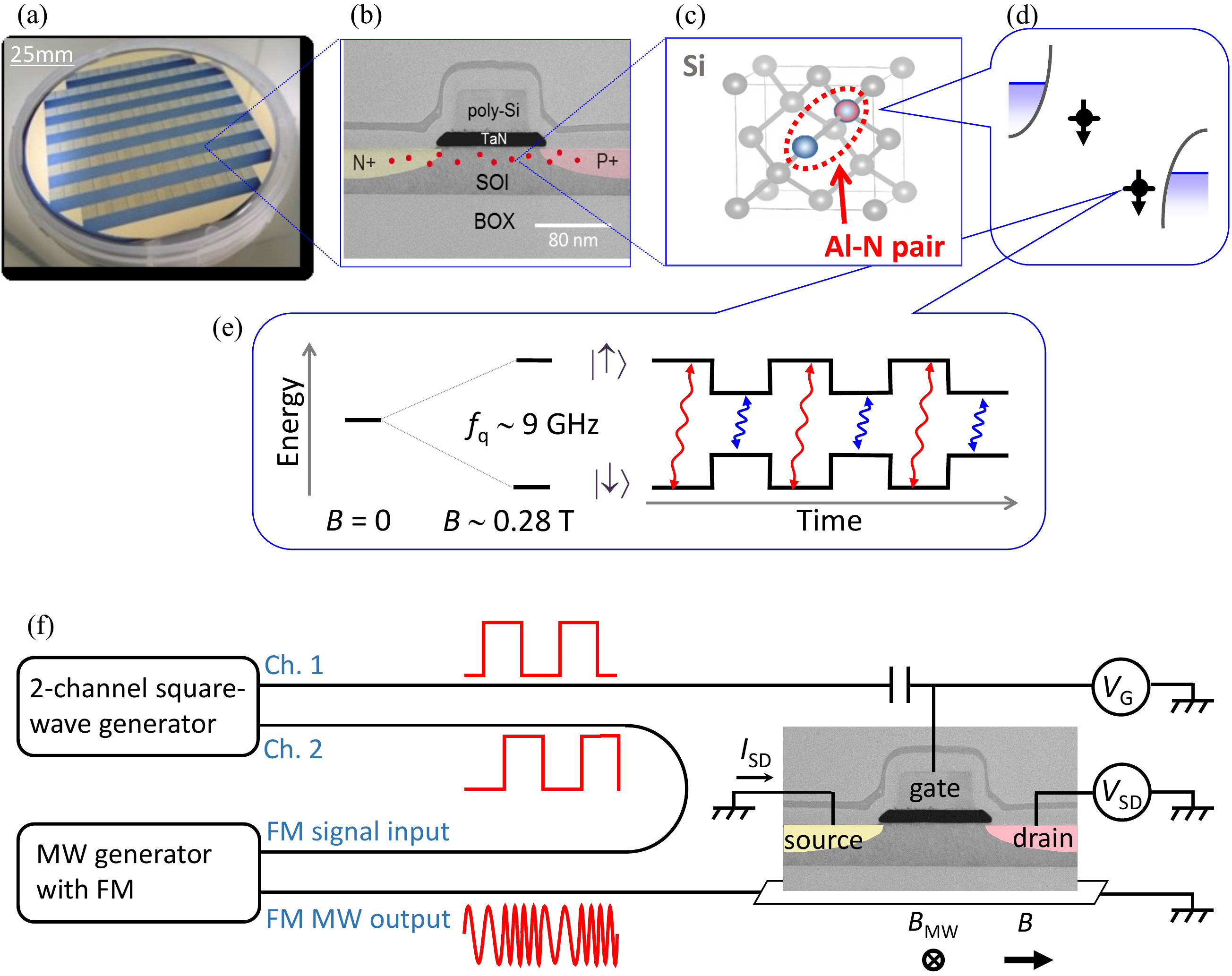}
\caption{Overview of device and its operation. The device is fabricated by
standard CMOS processing technology with ion-implantations of Al and N,
introducing Al-N impurity pairs as deep impurities. (a) Photo of a 4-inch
wafer after the ion-implantation processing. (b) Cross-section image by a
transmission electron microscope of the TFET-based device. (c) Schematic of
an Al-N impurity pair in the unit cell of Si. (d) Schematic diagram of the
potential energy landscape for PIN structure in the channel of the TFET with
two impurity levels. Electron tunneling via a deep impurity and a shallow
impurity exhibits both double-dot-like transport and spin blockade. (e)
Schematic of the time evolution of one of the spin qubits. The qubit energy
(Zeeman splitting) is modulated with a square-wave form $f_{\mathrm{q}}\pm 
\protect\delta \!f_{\mathrm{q}}$. The applied microwave frequency is also
modulated with a square-wave form $f_{\mathrm{MW}}\pm \protect\delta \!f_{%
\mathrm{MW}}$. (f) Schematic of the measurement set up. A two-channel
square-waveform generator and a MW generator with an FM signal are connected
for the double modulation. Schematic wave forms are illustrated as red
lines; a square wave from channel 1 (Ch.~1), a square wave from Ch.~2 that
has a defined phase difference against the channel~1 signal, and the FM MW
signal where its frequency is modulated by the channel~2 signal. }
\label{FigS1}
\end{figure}

\textit{TFET-based device.}--- A tunnel field effect transistor (TFET)
includes an $n$-type source electrode and a $p$-type drain electrode. Zener
tunneling carries on-current, which flows between a channel driven to the $p$%
-type (or $n$-type) electrode by a gate voltage and the $n$ electrode ($p$
electrode), see Fig.~\ref{FigS1}. This TFET-based device is attracting
attention as an ultra-low power consumption device because it enables
steeper switching than conventional metal-oxide-semiconductor field effect
transistors (MOSFETs). When a deep impurity is introduced into the channel
of a TFET having a short channel length, which can be regarded as a gated
PIN structure (with an undoped intrinsic semiconductor region between a $p$%
-type semiconductor and an $n$-type semiconductor region), a tunnel current
flows through the impurity level even at\ the off-state of the TFET.

Our TFET-based devices are fabricated in much the same process as
conventional MOSFETs. Using a $100$~nm thick Si-on-insulator (SOI) wafer
with $145$~nm thick buried oxide (BOX), the source and drain are formed by
shallow donor and acceptor ion implantation, respectively, and are activated
by high temperature rapid thermal annealing. Next, Al and N are
ion-implanted over the entire region consisting of source, channel, and
drain, and annealing is performed for a long time at a relatively low
temperature to form Al-N impurity pairs. A MOS gate is formed using TaN
high-k / metal and poly-Si.

The electrical conductivity of these devices was probed at a temperature of $%
1.6$~K. The source-drain current $I_{\mathrm{SD}}$ was measured while
applying the gate voltage $V_{\mathrm{G}}$ and the source-drain voltage $V_{%
\mathrm{SD}}$. A magnetic field $B$ was applied in parallel to the
source-drain current, and by applying a microwave current near the device,
an ac magnetic field was generated and used for electron spin resonance
(ESR).  For the synchronized modulation of both the gate voltage and the
microwave frequency, we used a two-channel square-waveform generator. The
two square-wave signals have the same frequency and tunable amplitudes and
phase difference. One of the square-wave signals is fed to the gate
electrode of the device via a DC block capacitor so that the gate voltage is
modulated by the square wave with an averaged DC voltage of $V_{\mathrm{G}}$%
. The other square-wave signal is fed to an auxiliary input as an analog
signal for the frequency modulation (FM) of the MW signal. 

\begin{figure}[t]
\begin{center}
\includegraphics[width=0.9\textwidth, keepaspectratio]{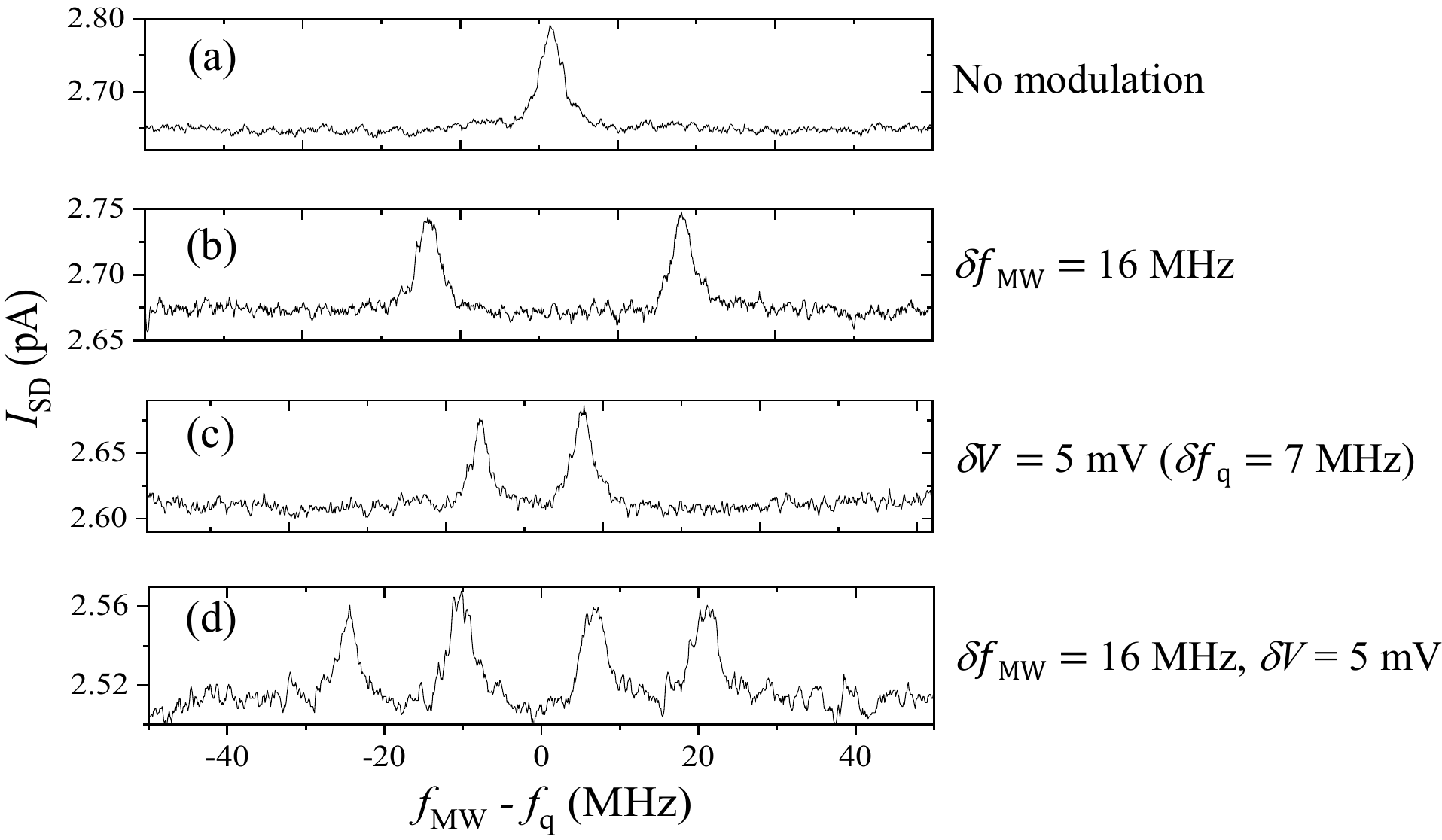}
\end{center}
\caption{Electron spin resonance (ESR) signal with and without modulation.
(a) without any modulation, one ESR peak (at $f_{\mathrm{MW}}=f_{\mathrm{q}%
}=9$~GHz) is observed in the source-drain current $I_{\mathrm{SD}}$ of the
device in the spin-blockade regime. Two ESR peaks for the modulation of the
microwave frequency (b) or for the modulation of the qubit energy (c). Panel
(d) presents four ESR peaks when the two modulations are applied
simultaneously. }
\label{FigS2}
\end{figure}

\textit{Without modulation.}--- The source-drain voltage $V_{\mathrm{SD}}$
and the gate voltage $V_{\mathrm{G}}$ of the device are set to satisfy the
spin-blockade condition, where the source-drain current shows an ESR
response. Figure~\ref{FigS2}(a) shows the ESR signal without any modulation,
with a single ESR peak observed where the microwave (MW) frequency $f_{%
\mathrm{MW}}$ matches the qubit frequency $f_{\mathrm{q}}\sim 9$~GHz. The
linewidth, or inverse coherence time $(T_{2}^{\ast })^{-1}=4$~MHz, is
limited by the lifetime of the spin-blockade state, see Ref.~[35] of the
main text.

\textit{Slow amplitude modulation.}--- We first simultaneously apply the
qubit energy modulation, with amplitude $2\delta \!f_{\mathrm{q}}$, and a
microwave frequency modulation (FM), with amplitude $2\delta \!f_{\mathrm{MW}%
}$. Both of these modulations have a square-wave form.

In order to realize the condition where the high/low frequency microwave
excites the large/small energy gap of the qubit, we tuned the parameters in
the following manner. The waveform used for the FM is a square wave with
frequency $0.05$~MHz, which was set to be much smaller than the inverse
coherence time $(T_{2}^{\ast })^{-1}$. The amplitude of the square wave was
set so that the microwave frequency is $f_{\mathrm{MW}}\pm \delta \!f_{%
\mathrm{MW}}$. Thus, if we sweep $f_{\mathrm{MW}}$, the two ESR peaks appear
at $f_{\mathrm{MW}}-f_{\mathrm{q}}=\pm \delta \!f_{\mathrm{MW}}$ [Fig.~\ref%
{FigS2}(b)].

Then we also turn-on the qubit energy modulation by adding a square-wave
modulation on $V_{\mathrm{G}}$. By increasing the modulation amplitude $%
\delta V_{\mathrm{G}}$ from zero, each of the two ESR peaks is further split
into two, and thus in total four peaks appear. Those four peaks correspond
to four combinations where the qubit with energy $f_{\mathrm{q}}\pm \delta
\!f_{\mathrm{q}}$ is excited by microwaves with frequency $f_{\mathrm{MW}%
}\pm \delta \!f_{\mathrm{MW}}$ [Fig.~\ref{FigS2}(d)]. Figure~2(b) of the
main text shows the evolution of the ESR peaks increasing $\delta V_{\mathrm{%
G}}$. The crossing of the lines of the ESR peaks appear at $\delta V_{%
\mathrm{G}}=11$~mV, where we see three peaks instead of four. The three
peaks correspond to these three situations:

(1) the qubit with frequency $\left( f_{\mathrm{q}}-\delta \!f_{\mathrm{q}%
}\right) $ is excited by microwaves with frequency $\left( f_{\mathrm{MW}%
}+\delta \!f_{\mathrm{MW}}\right) $;

(2) the qubit with $\left( f_{\mathrm{q}}-\delta \!f_{\mathrm{q}}\right) $
is excited by the microwave with $\left( f_{\mathrm{MW}}-\delta \!f_{\mathrm{%
MW}}\right) $ and, at the same time, the qubit with $\left( f_{\mathrm{q}%
}+\delta \!f_{\mathrm{q}}\right) $ is excited by the microwave with $\left(
f_{\mathrm{MW}}+\delta \!f_{\mathrm{MW}}\right) $; and

(3) the qubit with $\left( f_{\mathrm{q}}+\delta \!f_{\mathrm{q}}\right) $
is excited by the microwave with $\left( f_{\mathrm{MW}}-\delta \!f_{\mathrm{%
MW}}\right) $. Thus, at the center peak (peak at the crossing) $\delta \!f_{%
\mathrm{MW}}$ is matched to $\delta \!f_{\mathrm{q}}$. Now $\delta V_{%
\mathrm{G}}$ can be converted to $\delta \!f_{\mathrm{MW}}$, as shown in the
left axis of Fig.~2(b) of the main text.

\section{Theory of amplitude- and frequency-modulated single spin}

\subsection{Hamiltonian and master equation}

We describe our device as a modulated and driven quantum two-level system
with the pseudo-spin Hamiltonian%
\begin{equation}
H(t)=B_{z}(t)\sigma _{z}/2+B_{x}(t)\sigma _{x}/2.  \label{H}
\end{equation}%
\begin{table}[tbp]
\begin{tabularx}{\linewidth}{|>{\hsize=0.4\hsize}X|>{\hsize=1.8\hsize}X|>{\hsize=0.8\hsize}X|}
		\hline
		
		\centering Parameter 
		& \centering Description 
		& Value or Range \\ \hline
		
		\centering $ \omega _{\mathrm{q}}$ 
		& \centering $ =2\pi f_{\mathrm{q}} $, ESR frequency; this is defined by \newline the magnetic field, $\omega _{\mathrm{q}}\propto B$ 
		& \centering $ f_{\mathrm{q}}=9.0 $~GHz \newline at $B=0.28$~T \arraybackslash \\ \hline
		
		\centering $ \delta \omega _{\mathrm{q}}$ 
		& \centering $ =2\pi \delta \! f_{\mathrm{q}} $, amplitude of energy-level modulation; \newline  defined by the gate voltage modulation, $\delta \omega _{\mathrm{q}} \propto \delta V $ 
		& \centering $ \delta \! f_{\mathrm{q}} = 16 $ MHz \newline at $ \delta V = 12 $~mV \arraybackslash \\ \hline
		
		\centering $ \Omega $ 
		& \centering square-modulation frequency 
		& \centering $ \Omega \in [0.05,50] $ MHz$\cdot 2\pi $ \arraybackslash \\ \hline
		
		\centering $ \omega _{\mathrm{MW}} $ 
		& \centering $ =2\pi f _{\mathrm{MW}} $, driving MW frequency 
		& \centering $\omega _{\mathrm{MW}}=\omega _{\mathrm{q}}-\Delta \omega $,  \newline $\ \left\vert \Delta \omega\right\vert \leqslant 50$~MHz$ \cdot 2\pi $ \arraybackslash \\ \hline
		
		\centering $ \delta \omega _{\mathrm{MW}} $ 
		& \centering amplitude of frequency modulation 
		& \centering $\delta \omega _{\mathrm{MW}} = 32$ MHz$\cdot 2\pi $ \arraybackslash \\ \hline
		
		\centering $ G  $ 
		& \centering driving amplitude; this is defined by the microwave \newline power at the MW-generator output, $G\propto \sqrt{P_{\mathrm{MW}}}$ 
		& \centering $ G = 1$ MHz$\cdot 2\pi$ \newline at $ P_{\mathrm{MW}} = 7 $  dBm \arraybackslash \\ \hline
		
		\centering $ \phi $ 
		& \centering phase shift 
		& \centering $ \phi \in [0,360] $ degrees \arraybackslash \\ \hline
		
	\end{tabularx}
\caption{Description of the parameters controlling our driven and modulated
single-spin qubit}
\label{Table1}
\end{table}
To be more precise, we consider a single $1/2$-spin with a fast microwave
driving and a slow rf modulation of both amplitude and frequency. The
longitudinal part is defined by the Zeeman splitting, $B_{z}(t)=g(t)\mu _{%
\mathrm{B}}B$. The time-dependent gate voltage changes the $g$-factor by a
small value and we have 
\begin{equation}
B_{z}/\hbar =\omega _{\mathrm{q}}+\delta \omega _{\mathrm{q}}\cdot
s(t),\qquad \delta \omega _{\mathrm{q}}\ll \omega _{\mathrm{q}},
\end{equation}%
where $\omega _{\mathrm{q}}=2\pi f_{\mathrm{q}}$ represents the ESR
frequency, $\delta \omega _{\mathrm{q}}$ describes the \textit{amplitude
modulation}. In this work, we consider a square-wave modulation with the
signal%
\begin{equation}
s(t)=\mathrm{sgn}\left[ \cos \Omega t\right] .
\end{equation}%
The transverse part of the Hamiltonian is defined by the frequency-modulated
MW voltage applied to the substrate, 
\begin{eqnarray}
B_{x}/\hbar &=&2G\cos \left[ \omega _{\mathrm{FM}}(t)t\right] , \\
\omega _{\mathrm{FM}}(t) &=&\omega _{\mathrm{MW}}+\delta \omega _{\mathrm{MW}%
}\cdot s_{\phi }(t),  \notag \\
s_{\phi }(t) &=&\mathrm{sgn}\left[ \cos \left( \Omega t+\phi \right) \right]
,  \notag
\end{eqnarray}%
with amplitude $G$ and microwave circular frequency $\omega _{\mathrm{MW}%
}=2\pi f_{\mathrm{MW}}$. The modulation is assumed to be slow, i.e. $\Omega
\ll \omega _{\mathrm{MW}}$, and with a small amplitude, $\delta \omega _{%
\mathrm{MW}}\ll \omega _{\mathrm{MW}}$, where $\delta \omega _{\mathrm{MW}}$
describes the \textit{frequency modulation}. The theoretical and
experimental parameters are collected in Table~\ref{Table1}.

After the unitary transformation $U=\exp \left( -i\omega _{\mathrm{MW}%
}\sigma _{z}t/2\right) $ and the rotating-wave approximation, as in
Refs.~[35, 43, 45] of the main text, 
we obtain%
\begin{eqnarray}
H_{1} &=&\!\frac{\hbar }{2}\left[ \Delta \omega +\widetilde{f}(t)\right]
\sigma _{z}+\frac{\hbar G}{2}\sigma _{x},  \label{H_RWA} \\
\widetilde{f}(t) &=&\delta \omega _{\mathrm{q}}\!\cdot \!s(t)-\delta \omega
_{\mathrm{MW}}\cdot s_{\phi }(t),  \label{f}
\end{eqnarray}%
where $\Delta \omega =\omega _{\mathrm{q}}-\omega _{\mathrm{MW}}=2\pi \left(
f_{\mathrm{q}}-f_{\mathrm{MW}}\right) $.

Then, the upper-level occupation probability $P_{+}$ is readily obtained
from the stationary solution of the Bloch equations: 
\begin{equation}
P_{+}\!\!\left( \!\Delta \omega ,\frac{\delta \omega _{\mathrm{q}}}{\Omega }%
\!\right) \!=\frac{1}{2}\sum_{k=-\infty }^{\infty }\frac{G_{k}^{2}(\delta
\omega _{\mathrm{q}}/\Omega )}{G_{k}^{2}(\delta \omega _{\mathrm{q}}/\Omega
)\!+\!\frac{T_{2}}{T_{1}}\left( \Delta \omega \!-\!k\Omega \right) ^{2}\!+\!%
\frac{1}{T_{1}T_{2}}},  \label{Pup}
\end{equation}%
where $G_{k}(x)=G\left\vert \Delta _{k}(x)\right\vert $, which can be
interpreted as the dressed qubit gap, modulated by the function $\Delta
_{k}(x)$%
\begin{equation}
\Delta _{k}=\int\limits_{0}^{1}d\tau \exp \left[ -i2\pi k\tau
+i\int_{0}^{\tau }d\tau ^{\prime }\widetilde{f}\left( \tau ^{\prime }\right) %
\right] .  \label{delta}
\end{equation}%
The relaxation and decoherence times are denoted as $T_{1}$ and $T_{2}$,
respectively.

It is straightforward to calculate $\Delta _{k}$ using Eq.~(\ref{delta})
with the integrand $\widetilde{f}\left( \tau \right) $ defined by Eq.~(\ref%
{f}). The resulting expression for $\Delta _{k}$\ consists of four terms,
which have the denominators $-k\pm x\pm w$, respectively, with $x=\delta
\omega _{\mathrm{q}}/\Omega $ and $w=\delta \omega _{\mathrm{MW}}/\Omega $.
This defines the characteristic $k^{\prime }$s for each term. And,
respectively, the absolute values of $\Delta _{k}$\ for these four terms are
the following%
\begin{equation}
\left\vert \Delta _{k}\right\vert =\left\vert \frac{\sin \left[ \frac{\pi
-\phi }{2}\left( k\mp (x-w)\right) \right] }{\pi \left( k\mp (x-w)\right) }%
\right\vert ,\qquad \left\vert \frac{\sin \left[ \frac{\phi }{2}\left( k\mp
(x+w)\right) \right] }{\pi \left( k\mp (x+w)\right) }\right\vert .
\label{four}
\end{equation}

Consider now the limiting case $\Omega \rightarrow 0$, which describes the
incoherent-regime resonances at $\Omega <2T_{2}^{-1}$, as detailed in
Ref.~[35] of the main text. 
We then expect resonances along the lines%
\begin{equation}
\Delta \omega =k\Omega =\pm \delta \omega _{\mathrm{q}}\pm \delta \omega _{%
\mathrm{MW}}.
\end{equation}%
An analysis of the four relations from Eq.~(\ref{four}) in the limit $\Omega
\rightarrow 0$ gives for $\phi \in \lbrack 0,\pi ]$%
\begin{eqnarray}
\left\vert \Delta _{k}\right\vert &\simeq &\frac{\phi }{2\pi }\text{ \ \ \ \
\ \ \ for\ \ \ }k\simeq \pm \left( \delta \omega _{\mathrm{q}}+\delta \omega
_{\mathrm{MW}}\right) ,  \label{a} \\
\left\vert \Delta _{k}\right\vert &\simeq &\frac{\pi -\phi }{2\pi }\text{ \
\ \ for\ \ \ }k\simeq \pm \left( \delta \omega _{\mathrm{q}}-\delta \omega _{%
\mathrm{MW}}\right) .  \label{b}
\end{eqnarray}%
The above equations are useful for describing the experimental results, as
detailed in the next Section.

\subsection{Dressed states and heat-engine-like cycles}

\begin{figure}[t]
\begin{center}
\includegraphics[width=1 \textwidth, keepaspectratio]{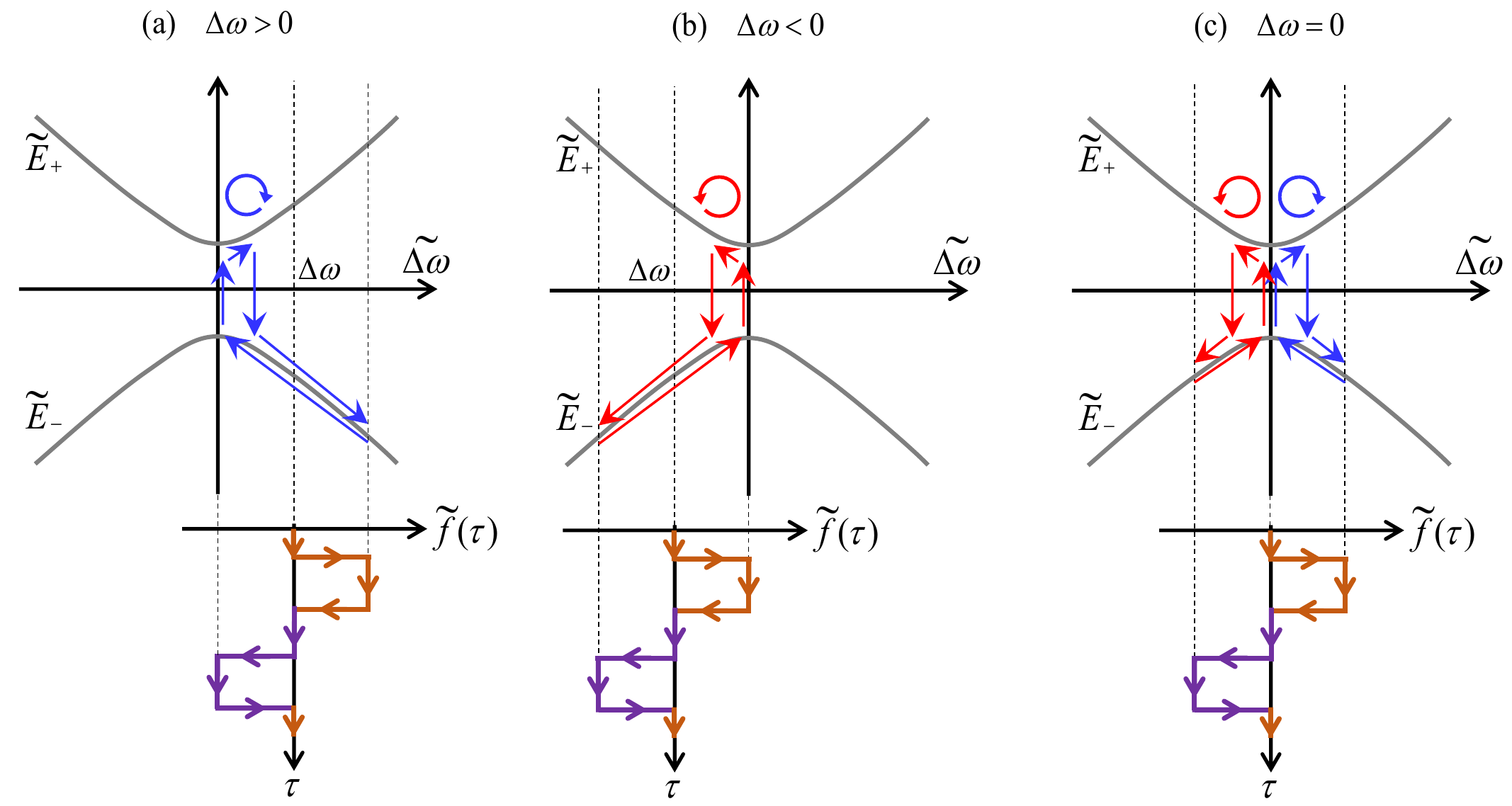}
\end{center}
\caption{Schematic of the refrigerator-like and engine-like cycles. The
dressed energy levels $\widetilde{E}_{\pm }$ are plotted as a function of
the modulation-defined frequency offset $\widetilde{\Delta \protect\omega }$%
. The modulation $\widetilde{f}(t)$ allows the system to oscillate between
two limiting values, with a possible excitation at the avoided-level
crossing and the subsequent relaxation. This is illustrated here for the
incoherent regime, $\Omega /2\protect\pi <2T_{2}^{-1}$. }
\label{Fig:cycles}
\end{figure}

Equation~(\ref{H_RWA}) describes the qubit dressed by the microwave signal.
The respective eigenstates are%
\begin{equation}
\widetilde{E}_{\pm }=\!\pm \frac{\hbar }{2}\sqrt{G^{2}+\widetilde{\Delta
\omega }^{2}},
\end{equation}%
where $\widetilde{\Delta \omega }=\Delta \omega +\widetilde{f}(t)$. In this
way, the time dependence in $\widetilde{f}(t)$ results in varying the
distance between the dressed-state energy levels $\widetilde{E}_{+}-%
\widetilde{E}_{-}$. The energy levels are plotted in Fig.~\ref{Fig:cycles}.
Then, we consider the dynamics forced by the modulation $\widetilde{f}(t)$,
around $\widetilde{\Delta \omega }=\Delta \omega $.

In figure~\ref{Fig:cycles}, we consider the symmetric case with $\delta
\omega _{\mathrm{q}}=\delta \omega _{\mathrm{MW}}$, which corresponds to the
horizontal dashed line in Fig.~2(b) of the main text. Then the modulation $%
\widetilde{f}(t)$ takes three values, $0$ and $\pm 2\delta \omega _{\mathrm{q%
}}$. We colour the curve for $\widetilde{f}(t)$ in brown when this
corresponds to the large-energy-gap half-period and in purple when this
corresponds to the small-energy-gap half-period.

Then we consider three situations in Fig.~\ref{Fig:cycles}(a-c), for $\Delta
\omega =2\delta \omega _{\mathrm{q}}$, $-2\delta \omega _{\mathrm{q}}\,$,
and $0$, respectively. In the situations (a) and (b), the qubit is excited
during the small-energy-gap and large-energy-gap stages, respectively. For
the incoherent regime, when $\Omega /2\pi <2T_{2}^{-1}$, the system relaxes
to the ground-state, and its evolution is shown by the blue and red
trajectories. In terms of Fig.~1 of the main text, these correspond to the
refrigerator-like and engine-like cycles. We mimic this by the blue and red
round arrows in Fig.~\ref{Fig:cycles}(a,b).

In the symmetric situation in Fig.~\ref{Fig:cycles}(c), there are two
cycles, both clockwise and counter-clockwise, during one modulation period.
We emphasize that there is no crosstalk between the two cycles, since the
relaxation is fast in the incoherent regime. In contrast, when increasing
the modulation frequency so that $\Omega /2\pi \gtrsim 2T_{2}^{-1}$, i.e.
going to the coherent regime, such cycles start to overlap. Essentially, 
\textit{such superposition of the two cycles results in the constructive or
destructive interference} with increased or suppressed upper-level
occupation probability. We consider these interference fringes in detail
both in the main text and below.

\section{Results}

\begin{figure}[t]
\begin{center}
\includegraphics[width=1\textwidth, keepaspectratio]{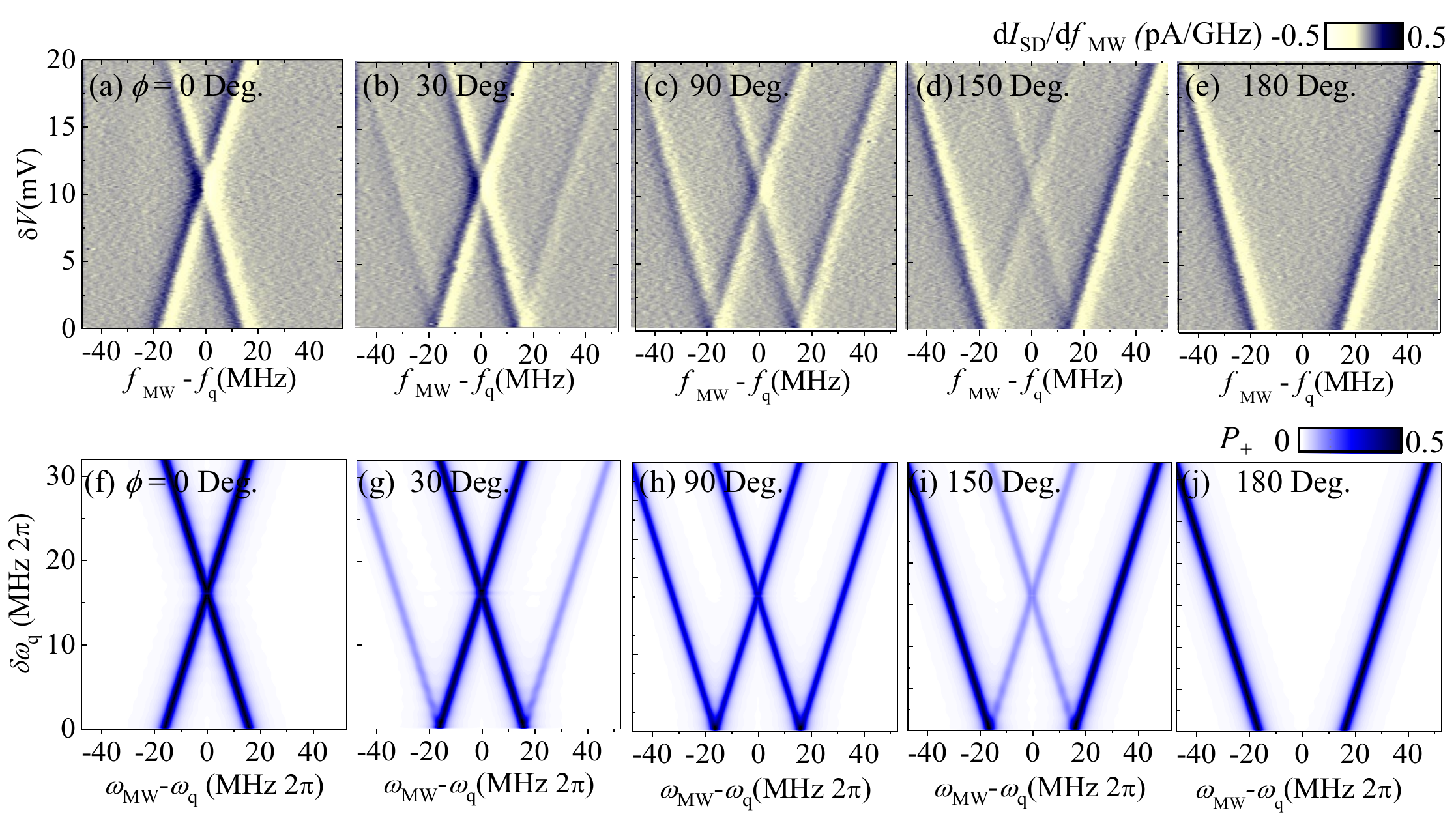}
\end{center}
\caption{The experimental source-drain current $I_{\mathrm{SD}}$, top row,
and the theoretical upper-level occupation probability $P_{+}$, bottom row,
as a function of the energy-level modulation amplitude $\protect\delta 
\protect\omega _{\mathrm{q}}=\protect\delta \protect\omega _{\mathrm{q}%
}\left( \protect\delta V\right) $ and the microwave frequency $\protect%
\omega _{\mathrm{MW}}=2\protect\pi f_{\mathrm{MW}}$. The data are similar to
the ones in Fig.~2(b) of the main text, for various phase shifts $\protect%
\phi $; the parameters are the same as in Fig.~2 of the main text, except of 
$\protect\phi $. The panels (a-e) and (f-j) show the experiment and
calculations, respectively. For $\protect\phi =0$ the in-phase condition is
met at $\protect\delta V$ $=12$~mV, where $\protect\delta \!f_{\mathrm{q}}$ $%
=\protect\delta \!f_{\mathrm{MW}}$, while the out-of-phase condition is met
at $\protect\delta V$ $=12$~mV and $\protect\phi =180$~deg. }
\label{FigS3}
\end{figure}

\begin{figure}[t]
\begin{center}
\includegraphics[width=0.9\textwidth, keepaspectratio]{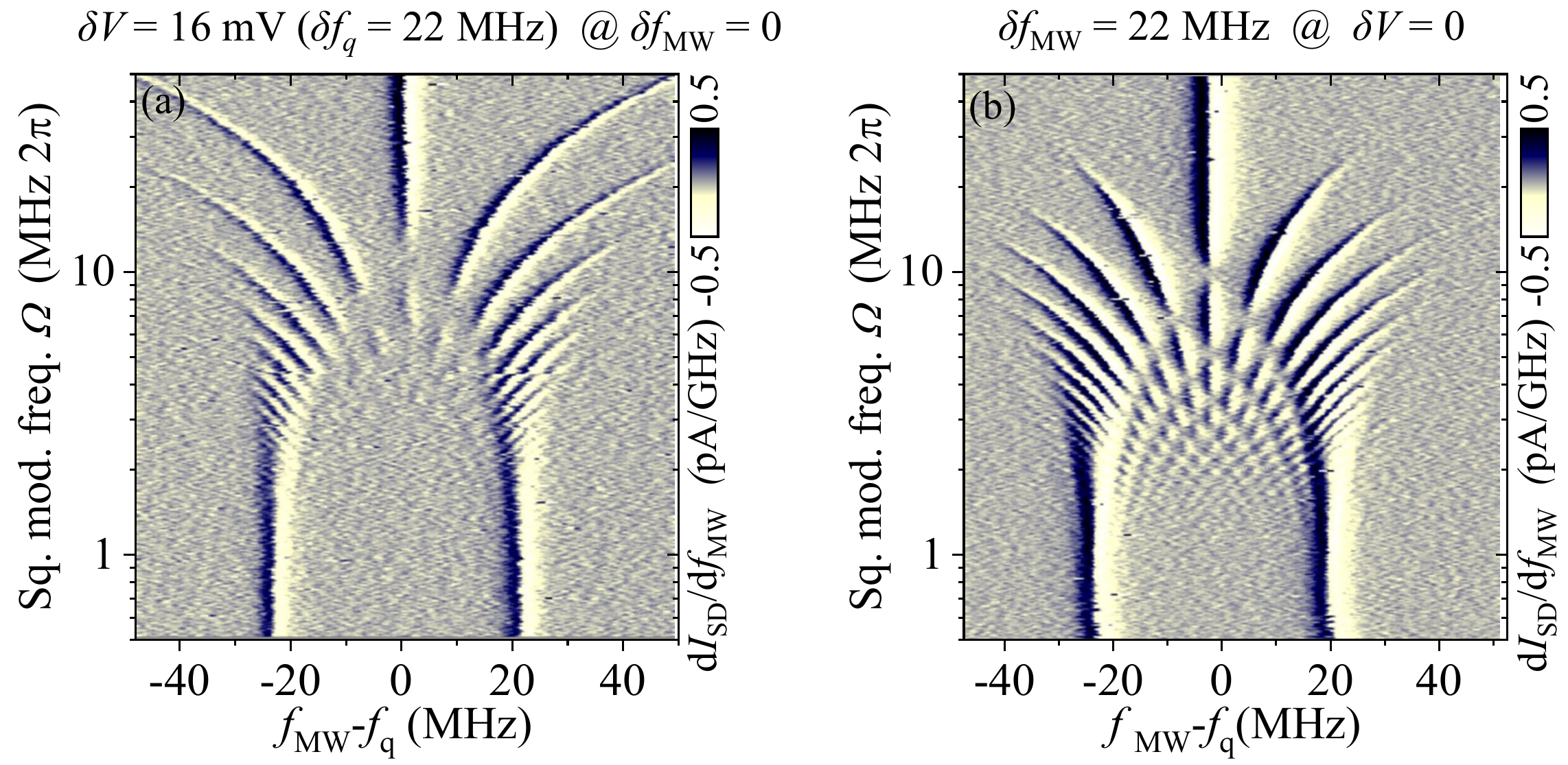}
\end{center}
\caption{Dependence of the source-drain current $I_{\mathrm{SD}}$ on the
square-wave modulation frequency $\Omega $~and the microwave frequency $f_{%
\mathrm{MW}}$ for $\protect\delta \!f_{\mathrm{MW}}=0$ (a) and $\protect%
\delta V=0$ (b). The amplitude of the modulation $\protect\delta V_{\mathrm{G%
}}$ is fixed to $8$~mV, which corresponds to $\protect\delta \!f_{\mathrm{q}%
}=22$~MHz. The interference pattern is the same as discussed in Ref.~[35] of
the main text and agrees with the calculations. This ensures that
square-wave signals up to $50$~MHz can be applied to the gate without
significant deformation of the waveform. In~(b) the FM frequency of the
microwave changes from $0.5$ to $50$~MHz. The amplitude $\protect\delta \!f_{%
\mathrm{MW}}$ of the FM is fixed to $22$~MHz. The interference pattern shows
a cut-off of the sideband peaks for modulation frequencies larger than $10$%
~MHz. Thus, we limit the FM modulation up to $5$~MHz in the following
measurements. }
\label{FigS4}
\end{figure}

\subsection{Experiments}

Measurements as in Fig.~2(b) in the main text, for various $\phi $, are
shown in Fig.~\ref{FigS3}(a-e). In Fig.~\ref{FigS3}(f-j) we demonstrate the
results of calculations related to the experimental ones in Fig.~\ref{FigS3}%
(a-e), respectively. Details of the calculations are described in the next
subsection.

The dependence of the source-drain current $I_{\mathrm{SD}}$ on the
square-wave modulation frequency $\Omega $~and the microwave frequency $f_{%
\mathrm{MW}}$ for $\delta f_{\mathrm{MW}}=0$ and $\delta V=0$ are shown in
Fig.~\ref{FigS4}(a) and Fig.~\ref{FigS4}(b), respectively.

\begin{figure}[t]
\begin{center}
\includegraphics[width=1\textwidth, keepaspectratio]{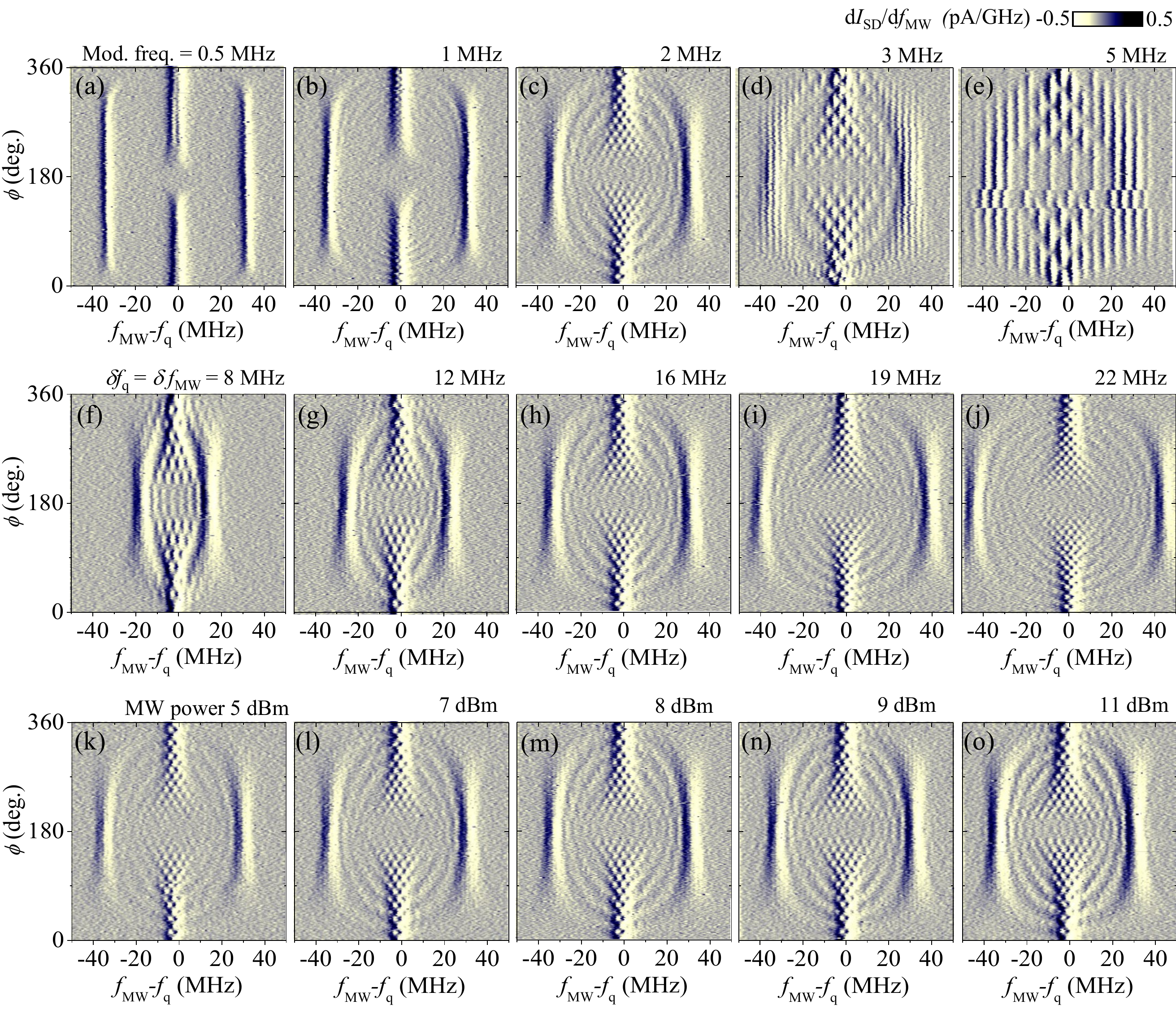}
\end{center}
\caption{Summary of the phase $\protect\phi $ dependence, as in Fig.~3 of
the main text, for various modulation frequencies $\Omega $~(a-e), various
modulation amplitudes $\protect\delta \!f_{\mathrm{MW}}$~(f-j), and various
microwave powers $P_{\mathrm{MW}}$~(k-o). Parameters are the same as in
Fig.~3, except the ones shown in this figure. For (f-j) similar measurements
as in~Fig.~2 were performed when $\protect\delta \!f_{\mathrm{q}}$ $=\protect%
\delta \!f_{\mathrm{MW}}$.}
\label{FigS5}
\end{figure}

Figure~\ref{FigS5} shows measurements as in Fig.~3(b) in the main text but
with different parameters, as written in each figure. We note that due to
the delay of the two modulation signals, the phase difference depends on the
modulation frequency within the MHz regime. Thus we redefine the $\phi =0$
position so that the intensity plot image becomes symmetric.

These experimental data are all reproduced by our calculations. Figure~\ref%
{FigS6} shows such calculated results, where the modulation-frequency
dependence corresponds to Fig.~\ref{FigS5}(a-e); see details in the next
subsection.

We extract the period of the notches from the data in Fig.~\ref{FigS5} and
summarize them in Fig.~\ref{FigS7}, showing excellent agreement between the
experiments and calculations.

Figure~\ref{FigS8}(b) shows one example of the measurements as in Fig.~\ref%
{FigS3}(a) ($\phi =0$) for the fast modulation frequency $\Omega /2\pi =2$%
~MHz. The X-shaped pattern in~\ref{FigS8}(b), similar to the one in Fig.~\ref%
{FigS3}(a), was obtained; but there are weak ripples around the pattern.
Note that if $\phi $ is set at a notch [$\phi =11$~deg., Fig.~\ref{FigS8}%
(c)], the X-shaped pattern shows a weak avoided crossing in the middle of
the green dashed horizontal line in Fig.~\ref{FigS8}(c).

\begin{figure}[t]
\begin{center}
\includegraphics[width=1\textwidth, keepaspectratio]{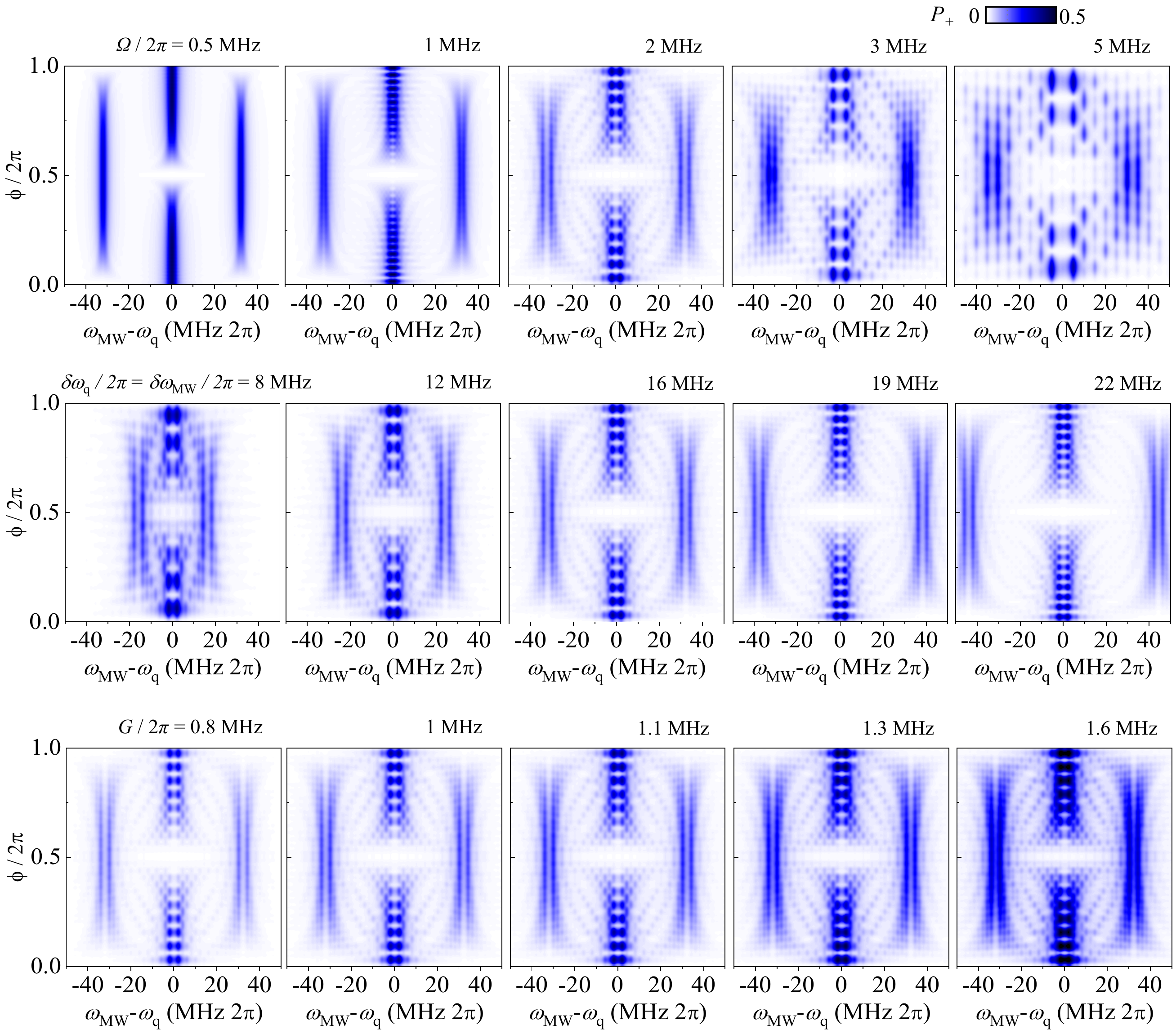}
\end{center}
\caption{Upper-level occupation probability $P_{+}$ as a function of the
frequency detuning $\left( \protect\omega _{\mathrm{MW}}-\protect\omega _{ 
\mathrm{q}}\right) $ and the phase difference $\protect\phi $, for several
values of the modulation frequency $\Omega $ (upper row), amplitude of
energy-level modulation $\protect\delta \protect\omega _{\mathrm{q}}$
(middle row), and driving amplitude $G$ (lower row).}
\label{FigS6}
\end{figure}

\subsection{Theory}

Using formulas (\ref{Pup}) and (\ref{four}) we can \textit{quantitatively}
describe the experimental results.

\subsubsection{Incoherent regime}

Consider, first, the incoherent regime, with $\Omega <2T_{2}^{-1}$. This is
described by the four possible resonance lines in Fig.~2 of the main text
and Fig.~\ref{FigS3}, where the lower row was plotted making use of Eqs.~(%
\ref{Pup}) and (\ref{four}). We note that the height and the width of the
lines are defined by the phase shift $\phi $.

In order to better understand the origin of the four lines in the W-shaped
pattern in Fig.~\ref{FigS3}, consider now Eqs.~(\ref{a}, \ref{b}). Then, at $%
\phi =0$, the width of the lines, described by Eq.~(\ref{a}), tends to zero,
and we have two lines only at 
\begin{equation}
\Delta \omega =\pm \left( \delta \omega _{\mathrm{q}}-\delta \omega _{%
\mathrm{MW}}\right) ,
\end{equation}%
which corresponds to the X-shaped lines in Fig.~\ref{FigS3}(f). They
intersect at $\delta \omega _{\mathrm{q}}=\delta \omega _{\mathrm{MW}}$,
which, importantly, can be useful for calibration of the power. Next, at $%
\phi =\pi $, the width of the lines, described by Eq.~(\ref{b}), tends to
zero, and we have two other lines, along 
\begin{equation}
\Delta \omega =\pm \left( \delta \omega _{\mathrm{q}}+\delta \omega _{%
\mathrm{MW}}\right) ,
\end{equation}%
which corresponds to the V-shaped lines in Fig.~\ref{FigS3}(j). For other
values of $\phi $, we have all the four lines. Their widths become equal for 
$\phi =\pi /2$, Fig.~\ref{FigS3}(h).

\subsubsection{Coherent regime}

To describe the coherent regime, with $\Omega \gtrsim 2T_{2}^{-1}$, as in
Fig.~\ref{FigS5}, we again use Eqs.~(\ref{Pup}) and (\ref{four}) to plot the
upper-level occupation probability $P_{+}$ in the general case, for the
respective parameters. Figure~\ref{FigS6} presents the appearance of the
interference fringes, where the upper-level occupation probability rapidly
changes between $0$ and $1/2$. This figure correctly reproduces the
experimental data in Fig.~\ref{FigS5}. Note that the upper panels of Fig.~%
\ref{FigS5} and Fig.~\ref{FigS6} demonstrate how the interference fringes
appear when increasing the modulation amplitude, starting from $\Omega $ $%
=0.5$~MHz$\cdot 2\pi <2T_{2}^{-1}$.

\begin{figure}[t]
\begin{center}
\includegraphics[width=0.6\textwidth, keepaspectratio]{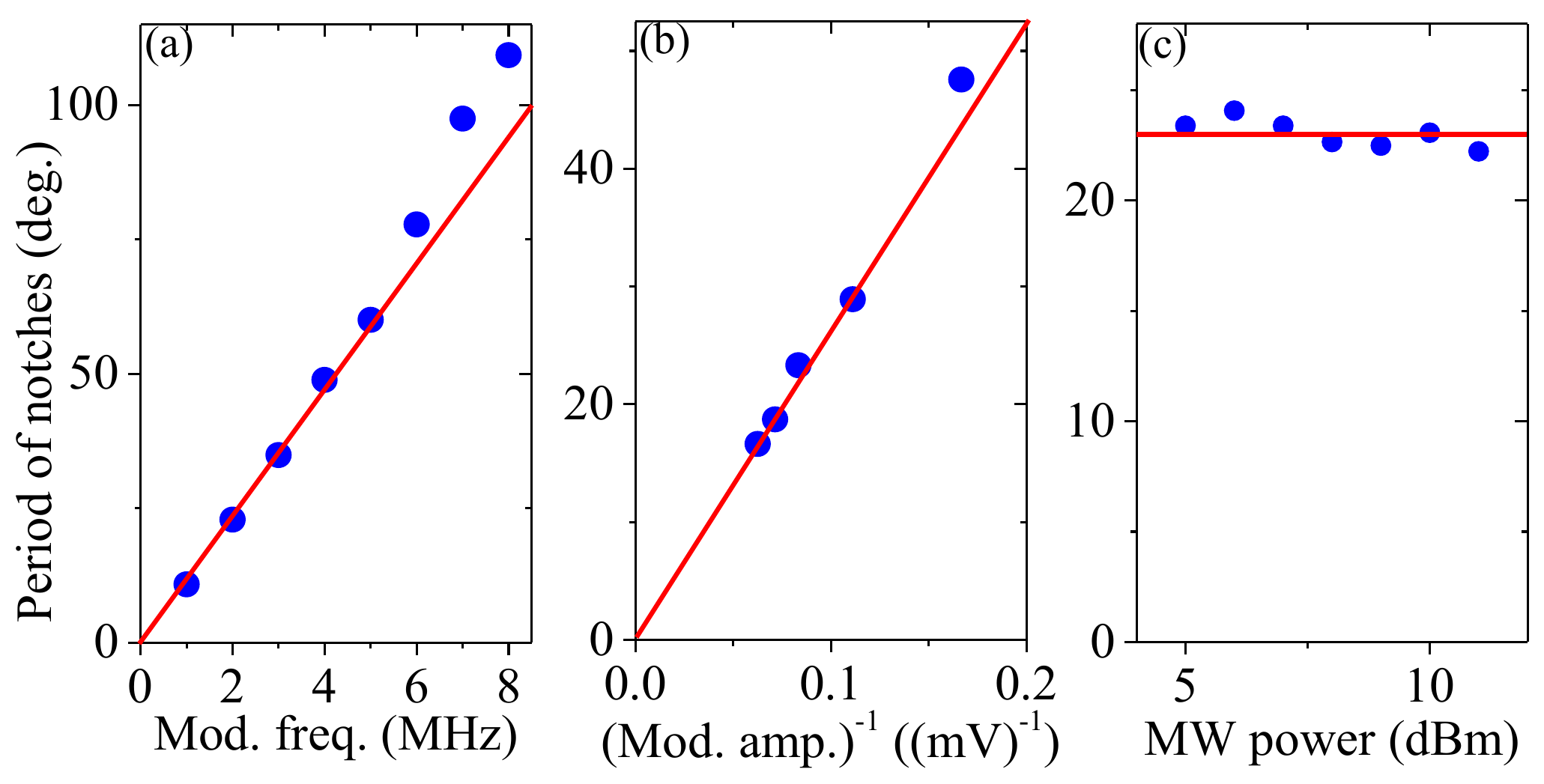}
\end{center}
\caption{Period of the notches (see Fig.~3 of the main text) as a function
of the modulation frequency $\Omega $, inverse modulation amplitude $\protect%
\delta V^{-1}$, and the microwave power $P_{\mathrm{MW}}$.}
\label{FigS7}
\end{figure}

We call \textquotedblleft notches\textquotedblright\ the suppression of the
resonance lines down to $P_{+}=0$, see Fig.~3 of the main text. From Eq.~(%
\ref{four}) it follows that the distance between the notches is defined by 
\begin{equation}
\Delta \phi /2\pi =\Omega /\delta \omega _{\mathrm{MW}}\propto \Omega \text{
\ \ \ \ and \ \ \ }\Delta \phi /2\pi =\Omega /\delta \omega _{\mathrm{q}%
}\propto \delta \omega _{\mathrm{q}}^{-1},  \label{notches}
\end{equation}%
if changing $\Omega $ and $\delta \omega _{\mathrm{q}}$, respectively.
Importantly, the distance between the notches is proportional to the
modulation frequency $\Omega $, inversely proportional to the modulation
amplitude $\delta \omega _{\mathrm{q}}$, and is independent of the driving
amplitude $G$, where the latter is defined by the MW power $P_{\mathrm{MW}}$%
. These relations are demonstrated in Fig.~\ref{FigS7}(a,~b, and~c),
respectively.

\begin{figure}[t]
\begin{center}
\includegraphics[width=0.88\textwidth, keepaspectratio]{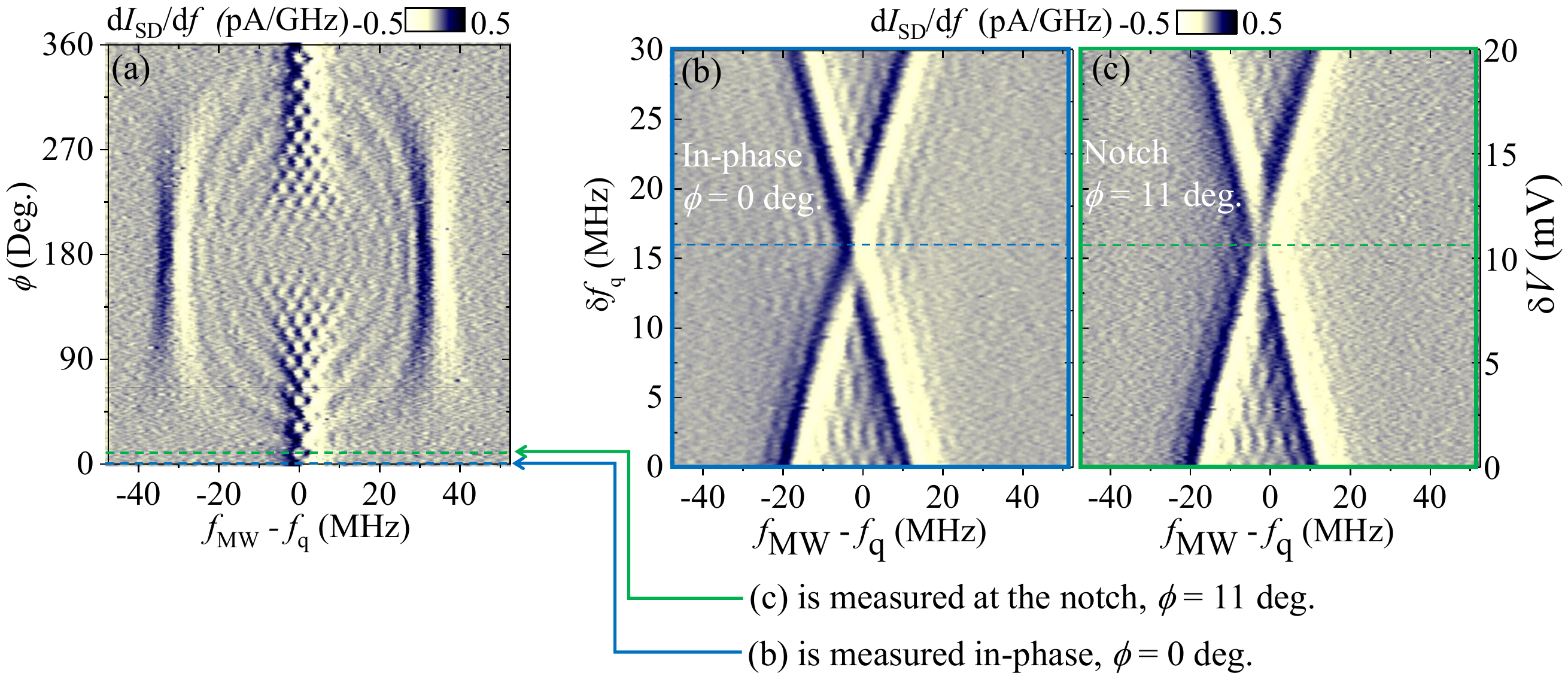}
\end{center}
\caption{Interference fringes in the dependence of the source-drain current $%
I_{\mathrm{SD}}$ for the X-shaped pattern. This is shown in (b) and (c) as a
function of the frequency detuning $\left( f_{\mathrm{MW}}-f_{\mathrm{q}
}\right) $ and the amplitude $\protect\delta \!f_{\mathrm{q}}$ of the
energy-level modulation, for two values of the phase difference $\protect%
\phi $, which correspond to the lines shown in (a).}
\label{FigS8}
\end{figure}

\subsubsection{Dynamics}

So far, we considered the stationary solution in order to describe the
experimentally observed data. The obvious success in this allows us to claim
that we have both the correct physical explanation of what happens and also
the correct parameters. Then, with this, we can go further and solve
numerically the Bloch equations to demonstrate the dynamics of the
upper-level occupation probability.

\begin{figure}[t]
\begin{center}
\includegraphics[width= 0.5\columnwidth, keepaspectratio]{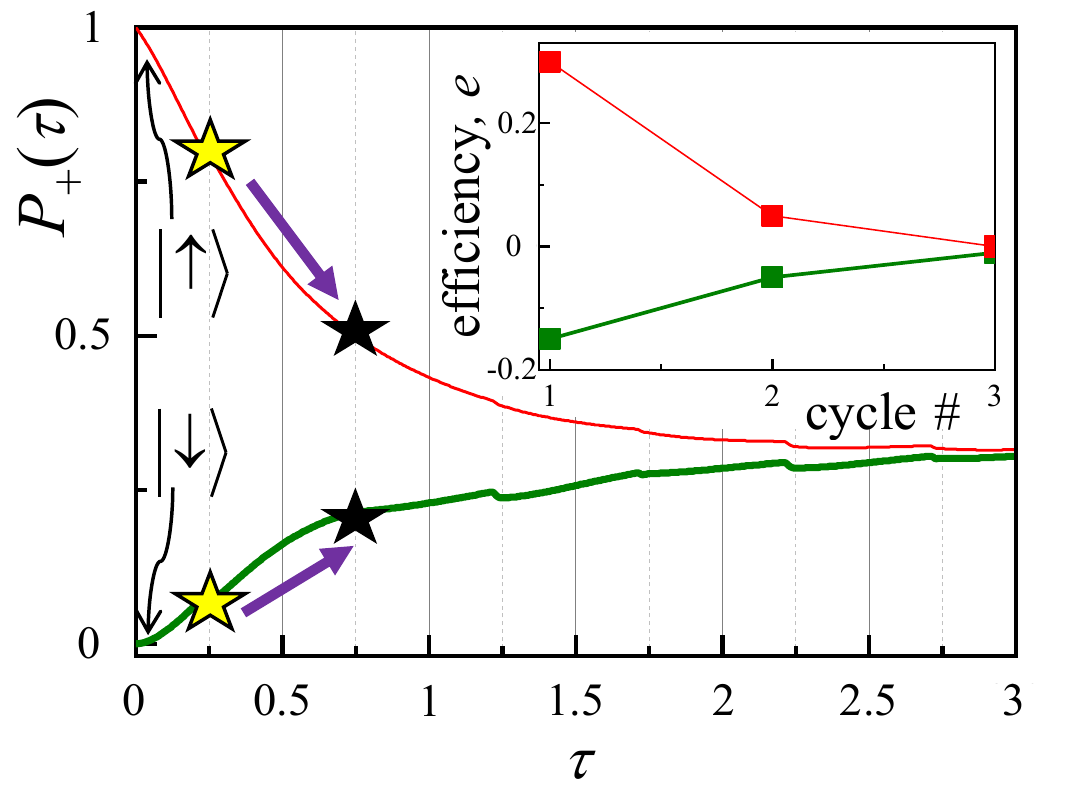}
\end{center}
\caption{Dynamics and the extended efficiency\textbf{\ }$e$\textbf{.}
Evolution of the upper-level occupation probability $P_{+}$, starting from
the ground state (thick green curve) and the excited state (thin red curve).
The inset shows the respective extended efficiency $e$, which is defined for
each cycle as the difference of the probability at the beginning and the end
of the small-energy-gap stage, shown by the yellow and black stars,
respectively.}
\label{Fig4}
\end{figure}

The time evolution of the upper-level occupation probability $P_{+}(t)$ for
the coherent case, with $\Omega /2\pi =2$~MHz, is shown in Fig.~\ref{Fig4}.
For this we solved the time-dependent Bloch equations numerically, with
different initial conditions. In a few cycles, the transient dynamics
evolves into the stationary solution. Note that Fig.~3 of the main text and
the graphs above show the time-averaged stationary solution, which
corresponds to the upper-level occupation probability $P_{+}$ at times $t\gg
T_{1,2}$. However, in principle, after preparing either the ground or
excited state, we can start up our device following one of the trajectories
in the main panel of Fig.~\ref{Fig4} and quantify it by the extended
efficiency $e$ in the inset. The extended efficiency is introduced there so
that $e=1$ corresponds to a perfect engine, as in Fig.~1(a) of the main
text, while $e=-1$ corresponds to a perfect refrigerator, as in Fig.~1(c).
Coexistence and interference of these two cycles gives $e\in (-1,1)$. A
decoherence or complete mixing of these cycles leads to $e=0$ and the system
does not work as a heat engine anymore.

In order to see the dynamics in the experiment, after initializing the
qubit, we would need a pulse-modulation of the drive frequency + qubit
energy. The pulse length is expected to be about 0.1--1~$\mu $s from the
previous Rabi oscillation experiment (Ref.~[28]). However, such experiments
have not been possible because the current experimental setup does not have
enough equipment to perform this pulsing. Thus, the experiments on the
dynamics are outside of the scope of this paper. In the future, we would
like to conduct the pulse experiments related to our calculated predictions,
shown in Fig.~S10. 

\begin{figure}[t]
\begin{center}
\includegraphics[width=1\textwidth, keepaspectratio]{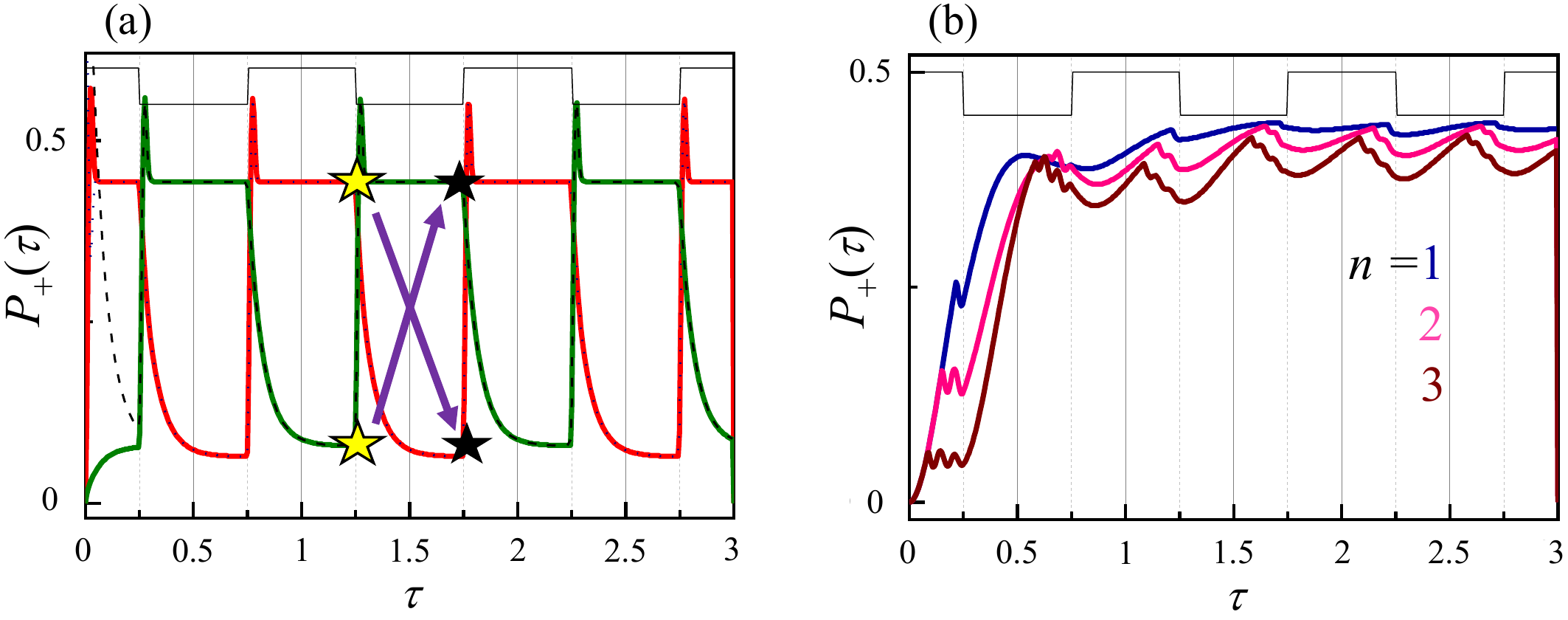}
\end{center}
\caption{Incoherent and coherent dynamics. The upper-level occupation
probability $P_{+}$ is plotted as a function of the dimensionless time $%
\protect\tau =\Omega t/2\protect\pi $. Thin solid lines are given as a guide
for the eye in order to mimic the large- and small-energy gap stages. The
square modulation frequency $\Omega /2\protect\pi $ is $0.2$~MHz for (a) and 
$2$~MHz for (b). }
\label{FigS0}
\end{figure}

Results of calculations are further presented in Fig.~\ref{FigS0} for (a)
the incoherent ($\Omega <2T_{2}^{-1}$) and (b) coherent ($\Omega \gtrsim
2T_{2}^{-1}$) cases. In~Fig.~\ref{FigS0}(a) the red and green curves
correspond to the resonant driving during either the high-energy stage or
the low-energy stage, which describe the upper or lower panels in Fig.~1 of
the main text, respectively. Thin dashed and dotted curves illustrate the
situation of starting from a different initial condition, which is being in
the excited state. Note that during the excitation, the system is first
resonantly excited and then relaxes to the stationary solution, which makes
a kink at the beginning of the \textquotedblleft ON\textquotedblright\
stage. In Fig.~\ref{FigS0}(b), the numbers $n=1,2,3$ denote the parameters,
taken for the first three resonant points in Fig.~3(e) of the main text
around $\delta \omega =0$ and counted from the bottom. At long times, $t\gg
T_{1,2}$, after the transient dynamics finishes, the upper-level occupation
probability tends to its stationary solution $P_{+}$. This is defined by the
competition of excitation and relaxation, and is given by Eq.~(\ref{Pup}),
which gives $P_{+}\leq 1/2$. 

\begin{figure}[t]
\begin{center}
\includegraphics[width=1\textwidth, keepaspectratio]{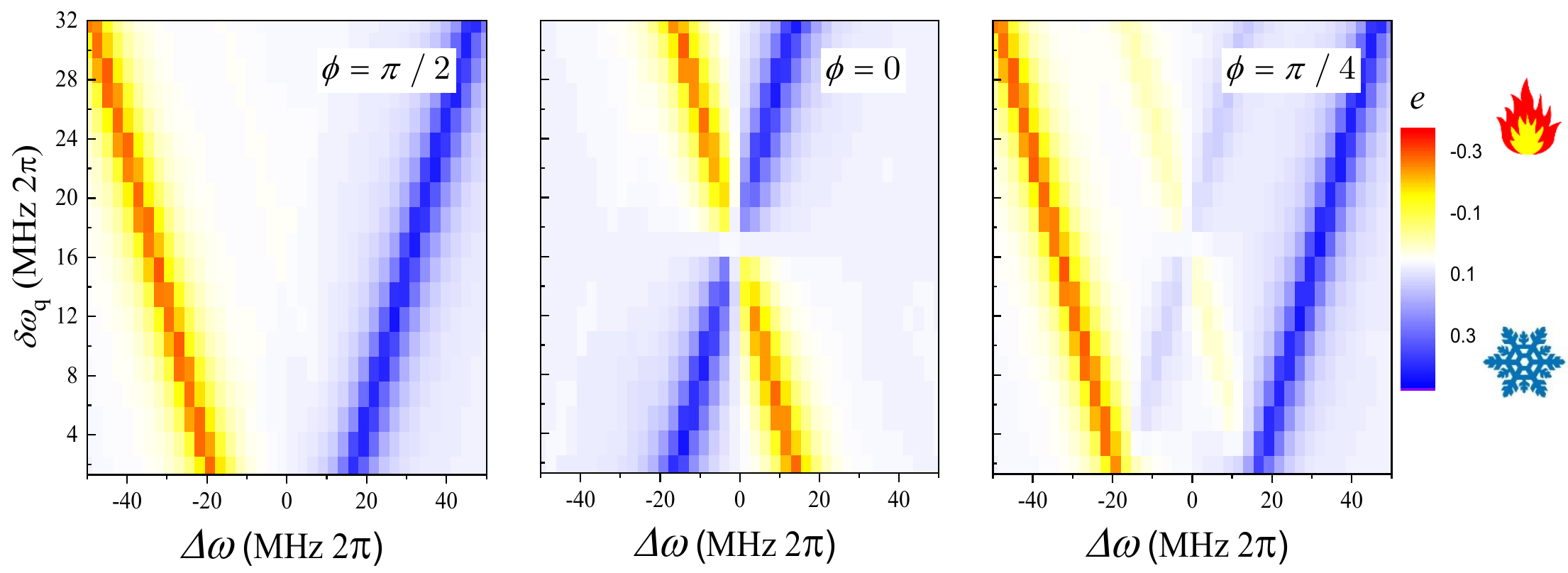}
\end{center}
\caption{Extended efficiency $e$ in the incoherent regime. This is
calculated for the same parameters as Fig.~\protect\ref{FigS3}, as a
function of the microwave frequency detuning $\Delta \protect\omega =\protect%
\omega _{\mathrm{MW}}-\protect\omega _{\mathrm{q}}$ and the energy-level
modulation amplitude $\protect\delta \protect\omega _{\mathrm{q}}$.}
\label{FigS10}
\end{figure}

In Fig.~\ref{FigS10} we present the calculated extended efficiency $e$,
which is defined for each cycle as the difference of the probability at the
beginning and the end of the small-energy-gap stage, see this in Fig.~\ref%
{FigS0}, shown by the arrows. Figure \ref{FigS10} is calculated for the
incoherent regime, for the same parameters as Fig.~\ref{FigS3}. Note that
for the stationary solution, the extended efficiency $e$ is close to $0$ in
the coherent regime (see Fig.~4 of the main text), while it displays both
the heat-engine and refrigerator regimes in the incoherent regime, as shown
in Fig.~\ref{FigS10} by the red and blue colours, respectively.

